\documentclass[epj]{svjour}
\usepackage{graphics}
\usepackage{epsfig}
\usepackage{amsbsy}

\begin{document}

\title{Final Results from phase II of the Mainz Neutrino Mass Search
       in Tritium $\boldsymbol \beta$ Decay}

\author{Ch. Kraus\inst{1,8}\thanks{This paper comprises principal parts
    of the PhD thesises of Christine Kraus, Beate Bornschein and Lutz
    Bornschein.} \and B.  Bornschein\inst{1,2a}\and L.
    Bornschein\inst{1,3a}\and J. Bonn\inst{1}\and B. Flatt\inst{1}\and A.
    Kovalik\inst{4}\and B. Ostrick\inst{1,6}\and E.W. Otten\inst{1}\thanks 
    {corresponding author: e-mail: Ernst.Otten@uni-mainz.de}\and 
    J.P.  Schall\inst{1}\and 
    Th. Th\"ummler\inst{1,6}\and Ch. Weinheimer\inst{1,5,7}}

  \institute{Institut f\"ur Physik der Johannes Gutenberg-Universit\"at
  Mainz, D-55099 Mainz, Germany \and present address: Forschungszentrum
  Karlsruhe, Tritiumlabor, D-76344 Eggenstein-Leopoldshafen, 
  Germany\and present address: Universit\"at Karlsruhe (TH), 
  Institut f\"ur exp. Kernphysik, Postfach 6980, D-76128 Karlsruhe, 
  Germany \and on leave from the Nuclear Physics Institute of the Acad. Sci. 
  Czech Republic, CZ-25068 Rez near Prague 
  \and Helmholtz-Institut f\"ur Strahlen und
  Kernphysik, Universit\"at Bonn, D-53115 Bonn, Germany
  \and present address: Helmholtz-Institut f\"ur Strahlen und
  Kernphysik, Universit\"at Bonn, D-53115 Bonn, Germany 
  \and present address: Institut f\"ur Kernphysik, Universit\"at M\"unster, 
  D-48149 M\"unster, Germany
  \and present address: departement of physics, Queen's university, 
  K7L3N6 Kingston, Canada}  

\titlerunning{Final Results from the Mainz Neutrino Mass Experiment}

\abstract{The paper reports on the improved Mainz experiment on tritum 
$\beta$ spectroscopy which yields a 10 times' higher signal to background
ratio than before. The main experimental effects and systematic uncertainties 
have been investigated in side experiments and possible error sources
have been eliminated. Extensive data taking took place in the years 1997 to 
2001. A residual analysis of the data sets yields for the square of the 
electron antineutrino mass the final result of 
$m^2(\nu_e)=(-0.6  \pm 2.2_{\rm{stat}} \pm 2.1_{\rm{syst}})$ eV$^2$/c$^4$. 
We derive an upper limit of $m(\nu_e)\leq 2.3$ eV/c$^2$ at 95\% confidence 
level for the mass itself.}


\maketitle

\section{Introduction}
\label{intro} In recent years observations of atmospheric, solar and reactor 
neutrinos \cite{SuperK,Davis,SK,original,Gallex,Sage,SNO,SNO2,Kamland}
in large underground detectors have discovered and established strong
mixing among the 3 neutrino generations $\nu_1$, $\nu_2$, $\nu_3$
produced in weak decays. The mixing manifests itself in
neutrino flavour oscillations whose wave numbers are proportional to
the differences of the squared masses ${\rm \Delta} m^2_{ij}=\mid m^2(\nu_i) 
-m^2(\nu_j) \mid$ of the mixing generations. Neutrino flavour
eigenstates $\nu_e$, $\nu_\mu$, $\nu_\tau$ produced in weak
interactions with electrons, muons or taus are thus connected to the
mass eigenstates $\nu_1$, $\nu_2$, $\nu_3$ through a unitary mixing
matrix $U$. So far oscillations $\nu_e \rightarrow \nu_\mu$ and
$\nu_\mu \rightarrow \nu_\tau$ have been observed yielding mass
differences $5.5 \cdot 10^{-5}$~eV$^2$/c$^4$ $\leq {\rm \Delta} m^2_{12} \leq
1.9 \cdot 10^{-4}$~eV$^2$/c$^4$ and $1.4 \cdot 10^{-3}$~eV$^2$/c$^4$ 
$\leq {\rm \Delta} m^2_{23} \leq 6.0 \cdot 10^{-3}$~eV$^2$/c$^4$, taken from a 
recent combined analysis of oscillation parameters \cite{Garcia}.

The fundamental discovery of finite mass differences between neutrino
generations has re-stimulated the question about their absolute scale
which is left open by any kind of interference experiment,
necessarily. It could range from a hierarchical ordering with $m^2_1$
or $m^2_3$ being much smaller than either of the measured ${\rm \Delta}
m^2_{ij}$ values to a quasi degenerate situation where these
differences are sitting on a much higher socket $m^2 \gg {\rm \Delta}
m^2_{ij}$ (see e.g. \cite{Garcia}). Assuming $m_1 \approx 0$ eV/c$^2$,
the former case would yield $m_2 \approx \sqrt{{\rm \Delta} m^2_{12}}
\approx 0.01$ eV/c$^2$ and $m_3 \approx \sqrt{{\rm \Delta} m^2_{23}}\approx
0.05$ eV/c$^2$.
An experimental hint towards a degenerate solution came recently from
a reanalysis \cite{Klapdor01,Klapdor04} of earlier, and from new data of 
the Heidelberg Moscow experiment on neutrinoless double $\beta$ decay of
$^{76}$Ge. If due to virtual emission and reabsorption of Majorana
neutrinos the observed rate would correspond to a so-called effective
neutrino mass
\begin{equation}
m_{ee} = \mid \sum m(\nu_j) \mid U_{ej} \mid ^2 e^{i \phi_j} \mid
\label{gln_effectiv}
\end{equation}
in the limits 0.1~eV/c$^2$ $\leq m_{ee} \leq 0.9$~eV/c$^2$ (99.7\%~C.L.)
\cite{Klapdor04}. 
The $\phi$ are phase factors of the mixing matrix $U$.
Although based on a 4${\rm \sigma}$ signal, this decay mode could still be 
modified by the exchange of some other non-standard particles.

Since in the universe a huge amount of about 336 relic
neutrinos/cm$^3$ are supposed to be left over from the Big Bang, a
sufficient rest mass could play an important role in the total mass
balance, in particular as so-called hot dark matter during the early
phase of cosmic evolution. Here the fine granulation of fluctuations,
observed in the temperature of the cosmic microwave background (CMB)
as well as in the large scale structure of the distribution of
galaxies (LSS), constrains the neutrino mass.  Combined analyses of
recent surveys yield upper mass limits of 0.23~eV/c$^2$ \cite{Spergel}
or somewhat more conservatively 0.33~eV/c$^2$ \cite{Hannestad,Barger}
assuming 3 degenerate neutrino generations.  Another analysis quotes a
finite mass of $\sim$0.2~eV/c$^2$ even \cite{Allen}. Still
there is a caveat in this kind of analysis: it results from fitting a
parametrized cosmological model in which 95\% of the gravitational
potential have to be attributed to unknown sources of matter and energy. 
\par From the above discussion we conclude that a model independent,
absolute mass measurement is indispensable even if the sensitivity
limit of alternative, model dependent methods is not reached yet. 
\par Among the model independent measurements, the investigation of the
$\beta$ spectrum of tritium near its endpoint has yielded by far the
most sensitive limits on the neutrino mass (strictly speaking the mass
of the electron antineutrino) in the past. Until the early nineties
magnetic $\beta$ spectrometers prevailed (reviewed in 
\cite{Robertson,Holzschuh}); thereafter electrostatic filters with magnetic
adiabatic collimation (MAC-E-Filters) took over thanks to their higher 
transmission and resolution (reviewed recently in 
\cite{WilkersonRobertson,Weinheim03}). They were proposed and realized 
independently in Mainz \cite{Picard} and Troitsk \cite{lobashev85,Belesev}.  
Our spectrometer yielded first results in 1991 from which we have extracted 
an upper limit of $m(\nu_e) < 7.2$~eV/c$^2$ (95\%~C.L.) \cite{Weinheim93}; 
it improved to 5.6~eV/c$^2$ at the end of phase I of this experiment 
\cite{Backe96}. These early results
still suffered from small spectral distortions farer off the endpoint
with a tendency to draw the observable $m^2(\nu_e)$ into the unphysical
negative sector the more, the farer the spectral interval, used in the
analysis, was extended below the endpoint. 

After the reason for this effect had been identified, we performed in
the years 1995-97 a substantial improvement program. It solved not
only that problem but also improved the signal to background ratio by
a factor of 10. In 1997 we started phase II of running, yielding in
the first year a limit down to 2.8~eV/c$^2$ (95\%~C.L.) \cite{Weinheim99} 
which was published in parallel to a 2.5~eV/c$^2$ limit (95\%~C.L.) from 
Troitsk \cite{lobashev99}. In the second year of data taking our value 
improved  to 2.2~eV/c$^2$~(95\%~C.L.), communicated in \cite{Bonn}.  
This limit was obtained from the experimental result 
$m^2(\nu_e) = (-1.6 \pm 2.5_{\rm stat} \pm 2.1_{\rm sys})$~eV$^2$/c$^4$.  
In the present paper we are giving a final report on this experiment, 
its analysis and its results. 

The paper is organized as follows: In section 2 we resume briefly the
principle of the experiment and discuss its sensitivity. In section 3
we describe the improvement program carried out for phase II. In
section 4 we report on the data taking periods in the years 1997-2001. 
The data are analyzed and discussed in section 5. In section 6
follows a discussion of the results. Conclusions and outlook
are given in section 7.

\section{$\boldsymbol \beta$-spectrum and neutrino mass measured by an 
integrating electrostatic filter.}
\label{experiment}

\subsection{${\boldsymbol \beta}$-spectrum in T$\bf _2$ decay}

Since we observe only the kinetic energy $E$ of the $\beta$ particle
we are measuring actually a sum of $\beta$ spectra, leading each with
probability $P_i$ to a final state of excitation energy $V_i$ of the
daughter and with probability $|U_{ej}|^2$ to a neutrino mass eigenstate
$m(\nu_j)$. Hence the differential decay rate (Fig.1) is
\begin{eqnarray}
\frac{\textrm{d}R}{\textrm{d}E}  &=&  N \frac{G_f^2}{2 \pi^3 \hbar^7 c^5}
\cos^2(\Theta_c) |M|^2 F(E,{\rm Z}+1)\cdot \nonumber \\
& &p(E+m_e {\rm c}^2)\sum_{ij} P_i(E_0-V_i-E)\cdot \nonumber \\
& &|U_{ej}|^2\sqrt{(E_0-V_i-E)^2 - m^2(\nu_j) {\rm c}^4}
\label{gln_rate}
\end{eqnarray}
Here $N$ is the number of mother nuclei, $G_f$ the universal Fermi
coupling constant, $\Theta_c$ the Cabibbo angle, $M$ the nuclear decay
matrix element, $F(E,{\rm Z}+1)$ the Fermi function, $p$ the electron
momentum, $m_e$ the electron mass and $E_0$ the $Q$ value of the T$_2$ decay
minus the recoil energy of the daughter. $E_0$ marks the endpoint of
the $\beta$ spectrum in case of zero neutrino mass. For the decay of
molecular T$_2$ to the groundstate of the daughter molecular ion ($^3$HeT)$^+$ 
one derives from the most precise direct determination of the mass 
difference $m({\rm T})-m(^3 {\rm He})$~=~(18590.1$\pm$1.7)~eV/c$^2$ 
\cite{endpoint} an endpoint energy of $E_0$=(18574.3$\pm$1.7)~eV 
\cite{Weinprom} by taking into account the effects through recoil energy and 
apparative effects\footnote{The apparative effects are a combination of 
electric potential depression, work functions from substrate and electode 
system and polarization shift. In the given references 
\cite{Weinprom,Weinheim93} the notation $E_0$ represents the difference in 
the electrostatic potential of the point the electron starts on the source 
and the point it crosses the analyzing plane, which we will later describe 
as $eU_0$ (see Table 1).}. This is in good agreement with \cite{Weinheim93}. 

With respect to the required energy resolution, this rather low endpoint 
favours the choice of tritium. Moreover, the minimal number of
electrons in the daughter molecule facilitates the precise calculation
of its excitation spectrum $(P_i, V_i)$ in $\beta$ decay.  Another advantage of
tritium decay is its superallowed character with a matrix element as
large as $M$= 5.55 \cite{Robertson}. This leads to a
reasonably short half life of 12.3~a and high specific activity of
about 3~MBq per cm$^2$ and monolayer from a frozen T$_2$ source, in use
here.

The Fermi function can be approximated by \cite{Simpson}
\begin{equation}
F= \frac{x}{1-\exp{(-x)}} \cdot (1.002037-0.001427 \cdot v_{\beta}/{\rm c})
\label{gln_fermi}
\end{equation}
with x=2$\pi ({\rm Z}+1) \alpha$ c/$v_{\beta}$, $\alpha$= fine structure 
constant, $v_{\beta}=$ velocity of the $\beta$-particle.
Radiative corrections to the $\beta$ spectrum are been applied 
\cite{Repco,gardner}. However, they are rather small within our present 
accuracy limits, they give rise to a shift of $m^2(\nu_e)$ of a 
few percent of our total systematic uncertainty.  
One may also raise the point whether contributions from right handed 
currents might lead to measurable spectral anomalies \cite{ste98}. 
We have checked that the present limits on the corresponding right handed 
boson mass \cite{PART} rule out a sizeable contribution within our present 
experimental accuracy \cite{CKraus}. The excitation spectrum ($P_i$,
$V_i$) of the daughter $(^3$HeT)$^+$ has first been calculated by
Kolos, followed by a number of refined numerical calculations, e.g.
\cite{Fackler}.  We are using here the most recent ones by Saenz et al.
\cite{Saenz}. The excitation spectrum is shown in Fig.~\ref{fig:figure2}. 
The first group concerns rotational and vibrational excitation of 
$(^3$HeT)$^+$ in its electronic ground state; it comprises a fraction of 
$P_{\rm g}$=57.4\% of the total rate. Its mean excitation energy is 1.73~eV 
for a $\beta$ energy close to the endpoint. The same
amount of recoil energy goes into the center of mass motion of the
molecule and is considered already in the $E_0$ value given above. In
solid T$_2$ the recoil may excite some phonons in addition.  But in
sudden approximation, which is quite valid here, the mean overall
recoil energy will even then -- for a $\beta$ energy close to the endpoint --
remain at 3.76~eV, which is the ballistic energy the decaying nucleus 
would receive in classical mechanics.

After this first so-called elastic group we observe an important gap
in the spectrum up to the first excited electronic state of
$(^3$HeT)$^+$ at 24~eV.  This gap could in principle be filled by a
$^3$He + T$^+$ continuum which starts at the dissociation energy of 4~eV. 
But dissociation on the cost of the $\beta$- energy is strongly
disfavoured in the Born Oppenheimer approximation. At 30~eV the first
electronic continuum opens up in which we observe still strong
resonances until complete ionisation is achieved in the second
continuum beyond 80~eV. 

In solid T$_2$ the sudden appearance of an additional nuclear charge
may also excite a neighbouring molecule. Kolos et al. \cite{Kolos} have
calculated the chance of this spectator excitation to be approximately
$5.9\%$ which is taken into account with some modification (see also
section \ref{uncer}).

\begin{figure}
 \centerline{\resizebox{0.45\textwidth}{!}
{\includegraphics{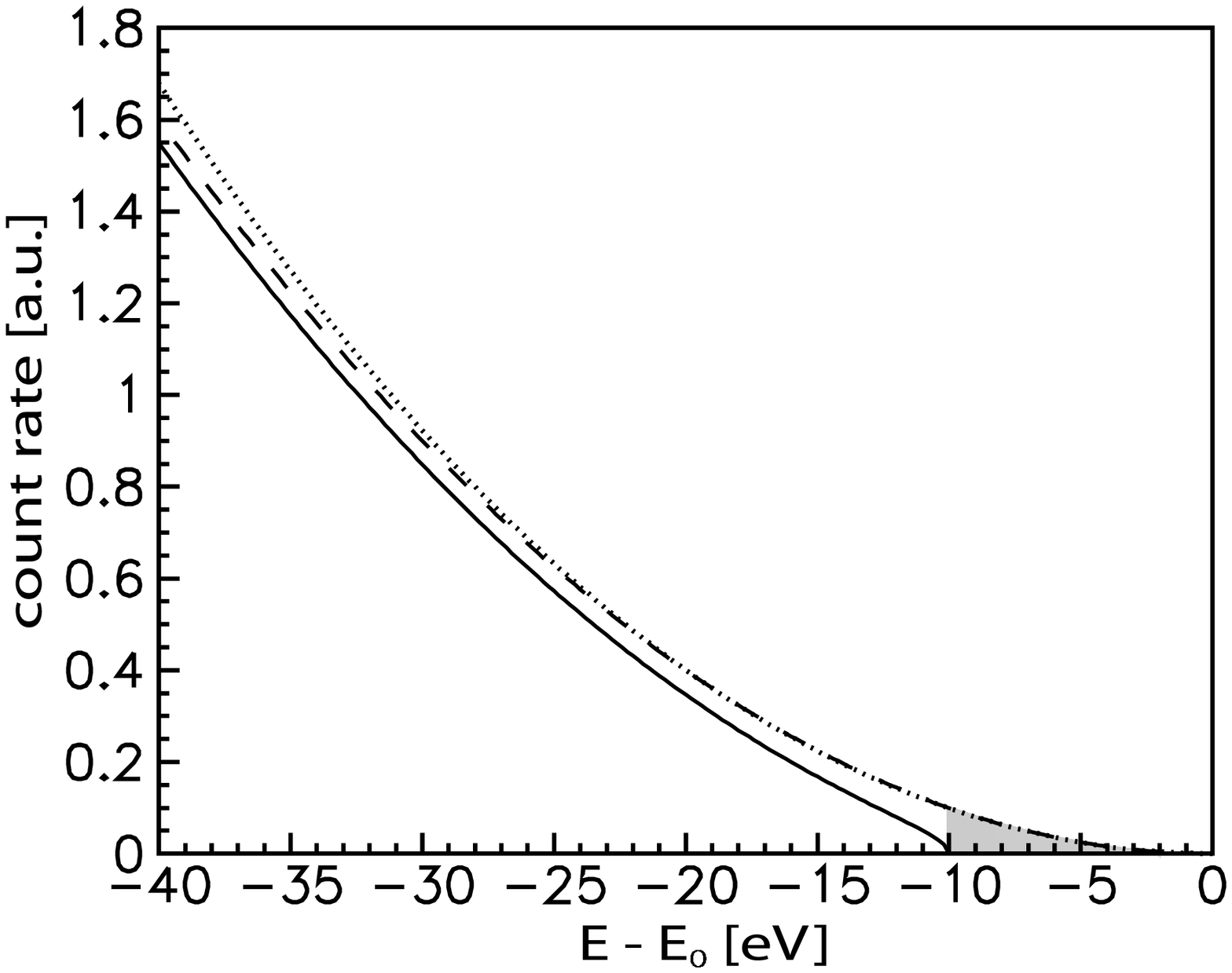}}}
\caption{Tritium $\beta$ spectrum close to the endpoint $E_0$. The dotted 
and the dashed line correspond to $m(\nu_e)=0$, the solid one to 
$m(\nu_e)=10$~eV/c$^2$. In case of the dashed and the solid line only the 
decay into the electronic ground state of the daughter is considered. 
For $m(\nu_e)$=10~eV/c$^2$ the missing decay rate in the last 10~eV below 
$E_0$ (shaded region) is a fraction of 2$\cdot10^{-10}$ of the total decay 
rate, scaling as $m^3(\nu_e)$.}
\label{fig:figure1}
\end{figure}

\subsection{Sensitivity of the ${\boldsymbol \beta}$-spectrum to  
${\boldsymbol m}^{\boldsymbol 2}({\boldsymbol \nu}_e)$}

\begin{figure}
\centerline{
\epsfig{file=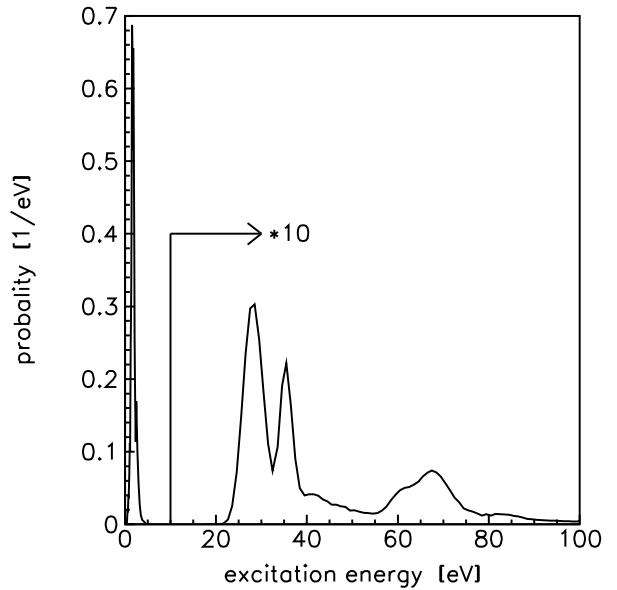, width=0.47\textwidth,angle=0}}
\caption{Excitation spectrum of the daugther 
($^3$HeT)$^+$ in $\beta$- decay of molecular tritium \cite{Saenz}.}
\label{fig:figure2}
\end{figure}

The last 2 terms in (\ref{gln_rate}) are the total energy $E_{\nu}$
and the momentum $p_{\nu}$ of the neutrino. They represent the
neutrino phase space and give rise to the parabolic increase of the
$\beta$ spectrum below $E_0$ for vanishing neutrino mass, shown in
Fig.~\ref{fig:figure1} by the dotted and dashed line. The solid line shows 
the effect of degenerate neutrino masses $m(\nu_j)=m(\nu_e)$= 10 eV/c$^2$. 
In case of the dashed and the solid line only the decay into the electronic 
ground state of the daughter is considered. For $m(\nu_e)$= 10 eV/c$^2$ the 
missing decay rate in the last 10 eV below $E_0$ is a fraction of 
2$\cdot10^{-10}$ of the total decay rate, scaling as $m^3(\nu_e)$.

We learn from these numbers that the tiny useful high energy end of
the spectrum is threatened by an enormous majority at lower energies.
However, it can be rejected safely by an electrostatic filter which
can be passed only by electrons with a kinetic energy $E$ larger than
a potential barrier $qU$ to be climbed. Any momentum analyzing, e.g.
magnetic spectrometer cannot guarantee this strict rejection since
scattering events may introduce tails to both sides of the resolution
function.

Actually, the $m(\nu_e)$ relevant signature of the spectrum extends further 
below the shaded triangle of missing count rate in (Fig.~\ref{fig:figure1}) 
into the region where $m(\nu_e)$ causes an asymptotically constant offset. 
Let us investigate this for a sharp filter which integrates the spectrum for 
energies $E>|qU|$. For short intervals we may treat all factors in front of 
the sum in (\ref{gln_rate}) as constant. In this interval it is sufficient 
to consider only the dominant decay mode into the electronic ground state 
(Fig.~\ref{fig:figure1}). We then obtain the integral count rate
\begin{eqnarray}
 R(E)  & = &\sum_j \int_{E}^{E_0-m(\nu_j) {\rm c}^2} 
            \frac{{\mathrm d}R}{{\mathrm d}E'} {\mathrm d}E'\nonumber\\ 
       & = C_R &\sum_j|U_{ej}|^2((E_0-E)^2- m^2(\nu_j) 
           {\rm c}^4)^\frac{3}{2}+b \nonumber\\
    & = & S + b
\label{gln_int}
\end{eqnarray}
where $b$ is the background rate, supposed to be independent of the
filter setting and $C_R$ is a specific signal rate. Under practical 
conditions the signal rate $S$ integrated over the measurement time $t$ 
separates from the background noise $\sqrt{bt}$ only at distances 
$E_0-E$ considerably larger than the sensitivity limit on the mass. There 
we may develop (\ref{gln_int}) to first order
\begin{equation}
R(E)= C_R((E_0-E)^3-\frac{3}{2}(E_0-E) \sum_j |U_{ej}|^2 m^2(\nu_j)
{\rm c}^4)+b.
\label{gln_intent}
\end{equation}
Besides the leading cubic term this approximate integral spectrum
displays a product of the interval length ($E_0-E$) and a weighted
squared mass
\begin{equation}
m^2(\nu_e)= \sum_j |U_{ej}|^2 m^2(\nu_j)
\label{gln_mass}
\end{equation}
which is our observable. Hence we call the square root
of (\ref{gln_mass}) the electron antineutrino mass $m(\nu_e)$ (see also
\cite{Weinheim03}).

The statistical noise $\sqrt{N}$ on the number of counts $N=(S+b)t$
after a measuring time $t$ will be dominated near $E_0$ by the
background and further below by the cubic term. The noise of the
latter rises like $(E_0-E)^\frac{3}{2}$ and hence faster than the
mass dependent signal. In between there must be a point with optimal
sensitivity on $m^2(\nu_e)$; it is found at
\begin{equation}
S=2b.
\label{gln_sb}
\end{equation}
From a measurement at that point for a time $t$ one would calculate 
\cite{Otten94} a statistical uncertainty by the help of (\ref{gln_intent})  
\begin{equation}
{\delta} m^2(\nu_e) {\rm c}^4= \left(\frac{16}{27}\right)^{1/6}
 \,C_R^{\frac{2}{3}} \,b^{\frac{1}{6}}\,t^{-\frac{1}{2}}.
\label{gln_deltam}
\end{equation}
We see that the dependence on the background rate is fortunately much
weaker than that on the specific signal rate. For the
characteristic parameters of our experiment $C_R=1.1 \cdot 10^{-5}$/eV$^3$s,
$b=0.015/$s, one finds the optimal point at 14~eV below $E_0$ and for
the value of (\ref{gln_deltam})
\begin{equation}
\delta m^2(\nu_e) {\rm c}^4 = 920 (t/\rm s)^{-\frac{1}{2}} {\rm eV}^2.
\label{gln_zahl}
\end{equation}
Within 10 days measuring time (\ref{gln_zahl}) drops to 1~eV$^2$. In an 
actual experiment one needs of course quite a number of measuring points 
within a reasonable interval in order to fix also the other parameters $C_R$, 
$E_0$, $b$ and to check the spectral shape in general by a $\chi^2$ fit.

At a particular measuring point $E$, an endpoint uncertainty $\delta E_0$ 
correlates to $\delta m(\nu_e)^2$ according to (\ref{gln_intent}) as
\begin{equation}
\delta m^2(\nu_e)= 
\frac{(\partial R/\partial E_0)}{(\partial R / \partial m^2(\nu_e))} 
\delta E_0= 2 (E_0-E) \delta E_0 /{\rm c}^4\,.
\label{gln_corr}
\end{equation}
Hence $\delta m^2(\nu_e)$ increases in proportion to the distance
from the endpoint, i.e. the neutrino energy $E_{\nu}$. This is the
crux of any missing mass experiment in relativistic kinematics where 
(\ref{gln_corr}) follows quite generally from the quadratic mass
energy relation $m^2 {\rm c}^4= E^2 -p^2 {\rm c}^2$. That underlines again 
the necessity of measuring the neutrino mass close to the $\beta$ endpoint 
and disfavours any other experimental concept involving
energetic neutrinos in order to gain phase space, i.e. rate.

Instead of fitting $E_0$ together with the other parameters from the
data one could consider to use the known $Q$ value instead \cite{endpoint}. 
Its error of 1.7~eV, however, would cause through (\ref{gln_corr}) in the
most sensitive region, i.e. around 14~eV below the endpoint, an error
in $m^2(\nu_e)$ of about 50~eV$^2$/c$^4$. This is far beyond our present 
value obtained from an inclusive fit. The latter is sensitive only to the 
easily measured small voltage differences in the scan rather than to the 
absolute energy scale.

On the other hand, we learn from (\ref{gln_corr}) that $E_0$ should be
fitted including somewhat larger distances from $E_0$, since its
uncertainty $\delta E_0$ decorrelates from $\delta m^2(\nu_e)$ like
1/$(E_0 - E)$. Altogether, there are in principle three spectral
regions from which the basic parameters: $b$, $m^2(\nu_e), E_0$ are
fitted most sensitively and with a minimum of crosstalk: \\
(i) ~~a region beyond $E_0$ fixing $b$, \\
(ii) ~a region shortly below $E_0$ fixing $m^2(\nu_e)$ and \\
(iii) a region further below $E_0$ fixing $E_0$. \\
In the latter region, however, the inelastic components of the spectrum
and their uncertainties start to matter which finally dominate the
systematic error. Hence we expect an optimal length of the measuring
interval at which we meet a proper balance between the systematic und
statistical uncertainty of the result.

\section{Improvements of the Mainz MAC-E-Filter}

\begin{figure}
  \centerline{\resizebox{0.5\textwidth}{!}{\includegraphics{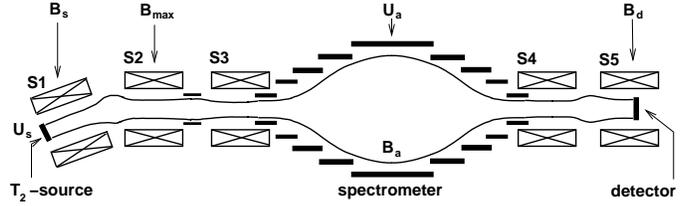} }}
\caption{The improved Mainz MAC-E-Filter is shown schematically. 
  The distance between source and detector is about 6~m and the
  diameter of the spectrometer vessel is 1~m. From left to right: 
  Frozen T$_2$ source housed in the tilted solenoid S1; guiding solenoids S2, 
  S3; the vessel with altogether 27 electrodes;
  refocussing solenoid S4, S5 housing the detector. The shown magnetic
  field lines confine the flux tube within which the $\beta$ particles are
  guided.}
\label{fig:figure3}
\end{figure}

$\beta$-spectroscopy in the endpoint region by an electrostatic
filter is particularly advantageous in combination with an electron optics 
based on the principle of magnetic adiabatic collimation (MAC-E-Filter) 
\cite{Picard,lobashev85}. Particles of charge $q$ are transported 
from the source to the detector by spiraling along the lines of a magnetic 
field $B$ connecting both (Fig.~\ref{fig:figure3}). Hence they can be accepted
in the full forward solid angle of 2$\pi$, in principle. An electrostatic 
filter potential $U$ in between is passed if the longitudinal energy 
$E_{\parallel}$ along the guiding
$B$-line is larger than $qU$. In order to filter the full energy
sharply the particle momenta have to be well collimated along
$B$. This is achieved by lowering the field strength from
a very high value $B_{\rm max}$ to a quite small one $B_{\rm a}$ in the region
of the analyzing potential. Thereby the transverse energy in the
cyclotron motion $E_{\bot}$ is reduced adiabatically in proportion to
the magnetic field strength and transformed into longitudinal energy
$E_{\parallel}$ along $B$.
The non relativistic limit the transformation reads:
\begin{equation}
\frac{E_{\bot \rm a}}{E_{\bot \rm max}}= \frac{B_{\rm a}}{B_{\rm max}} = 
\frac{{\rm \Delta} U}{U}.
\label{gln_res}
\end{equation} 
(\ref{gln_res}) defines the relative width ${\rm \Delta} U/U$ of a
MAC-E-Filter. If the field maximum is placed at the source, 
$B_{\rm max}=B_{\rm s}$, then it accepts the full forward solid angle 
${\rm \Delta} \Omega = 2 \pi$.

In reality we have limited the acceptance to a maximum start angle
$\Theta_{\rm max}$ by placing a field maximum $B_{\rm max} > B_{\rm s}$ in
between source and analyzing plane acting as magnetic mirror for
particles starting at angles $\Theta > \Theta_{\rm max}$ with
\begin{equation}
\Theta_{\rm max}={\rm \arcsin} \sqrt{B_{\rm s}/B_{\rm max}}.
\label{gln_angle}
\end{equation}
Moreover, the angular distribution is slightly modified from isotropy
by a scanning potential $U_{\rm s}$ on the source. Still the transmission
function is analytic \cite{Picard92}. For charges $q$=$-e$ it is given in 
the 4 adjacent intervals 

\begin{eqnarray*}
(i)  &:& E-e U_{\rm s} \leq-e U_{\rm a}\\
(ii) &:& -e U_{\rm a}<E-e U_{\rm s}<-e\cdot 
         U_{\rm a}~B_{\rm max}/(B_{\rm max}-B_{\rm a})\\
(iii)&:& -eU_{\rm a} B_{\rm max}/(B_{\rm max}-B_{\rm a}) \leq E -eU_{\rm s} 
         \leq E B_{\rm max}/B_{\rm s}\\
(iv) &:& (E-eU_{\rm s})/E \geq B_{\rm max}/B_{\rm s}
\label{bereiche}
\end{eqnarray*}

by

\begin{eqnarray}
T = \left\{%
\begin{array}{l@{\quad \mbox{}\quad}l}
  0 ~~~~~~~~~~~~~~~~~~~~~~~~~~~~~~~~~~~~~~\rm(i) \\ 
  1-\sqrt{1-\frac{E -e U_\mathrm{s} +e U_{\rm a}}{E}
       \frac{B_\mathrm{s}}{B_\mathrm{a}}} ~~~~ ~~~~(ii) \\
  1-\sqrt{1-\frac{E -e U_\mathrm{s}}{E} 
  \frac{B_\mathrm{s}}{B_\mathrm{a}} } ~~~~~~~~~~~~~~(iii)\\
  1~~~~~~~~~~~~~~~~~~~~~~~~~~~~~~~~~~~~~~(iv). \\  
\label{gln_trans}
\end{array} \right.
\end{eqnarray}

The second line of (\ref{gln_trans}) describes the sharp rise of
the transmission from 0 to a plateau within the filter width ${\rm \Delta}
U_{\rm s}={\rm \Delta} U$. The third line describes a further, slow rise of
the transmission in the plateau region as function of an accelerating,
thus forward focussing scanning potential $U_{\rm s}$ on the source until the 
mirror function of $B_{\rm max}$ is ruled out in~$(iv)$.
Transmitted electrons are refocussed by solenoid S4 and hit a
silicon detector in the centre of another solenoid S5 at a reduced
field strength $B_{\rm d}=0.31 \cdot B_{\rm max}$. This limits the angle of 
incidence to $34^\circ$. By the help of auxiliary coils around the central 
part of the spectrometer, $B_{\rm a}$ can be varied independently and hence 
the resolution through (\ref{gln_res}, \ref{gln_trans}). A ratio
\begin{equation}
B_{\rm a}/B_{\rm s} > A_{\rm s}/A_{\rm a}
\label{gln_auf}
\end{equation}
has to be observed, however, in order to keep the cross section of the
beam carrying flux in the analyzing plane well inside the cross
section $A_{\rm a}$ of the cylindrical electrodes. $A_{\rm s}$= 2 cm$^2$ 
is the cross section of the T$_2$ source.
We have been running at field ratios down to $B_{\rm a}/B_{\rm s} = 
3.3 \cdot 10^{-4}$,
which limits the flux tube diameter to 88~cm as compared to the
diameter of 94~cm of the central electrode.

These relations play a role for the background since the electrodes
will emit secondary electrons when they are hit by cosmic rays or any 
other particles originating from radioactivity. If accelerated toward the 
detector these electrons will arrive with an energy close to that of the 
transmitted $\beta$-particles and cannot be discriminated by the 1.4~keV 
(FWHM) resolution of the detector. 
It is important, therefore, that these secondary electrons are being
guided adiabatically along magnetic field lines which pass by the
detector. Still we observe enhanced background on its outermost ringsegments. 
Moreover the central guiding field should not be lowered below 
$B_{\rm a} \approx 5 \cdot 10^{-4}$~T in order to guarantee full transmission 
of the 200~eV energy interval under study \cite{Otten94}.
Another set of correction coils around the spectrometer annuls the
transverse component of the earth's magnetic field and steers the
$\beta$ flux.  Runs were performed at settings $B_{\rm max}=2.211$~T,
$B_{\rm a}=5.67 \cdot 10^{-4}$~T, $B_{\rm s}=1.087$~T 
($\Theta_{\rm max}=44.5^{\circ}$) or
$B_{\rm s}=1.693$~T ($\Theta_{\rm max}=61.6^{\circ}$), $U_{\rm a}=-18690$~V,
-20~V$\geq U_{\rm s}\geq$-320~V. More details on the general setup and 
function of the Mainz MAC-E-Filter have been given in \cite{Picard,Picard92},
and on its recent improvements and performance in \cite{BornB,BornL,Kraus}.

\subsection{The new source section} In the following we will focus
on the various improvements of the apparatus, performed in the years
1995-97 \cite{BornB}. A decisive improvement concerns the replacement
of the LHe bath cryostat by a flow cryostat which allowed to cool down
the T$_2$-carrying substrate below 2~K by a horizontal cooling
section, designed and built by Oxford Instruments on customer's
demand. Below that temperature the shock condensed, amorphous T$_2$-films
have been proven to be stable in time. Earlier the source had been
operated at temperatures between 3~K and 4~K, at which these films
turned out to dewet from the substrate and to contract into small
crystals with an average thickness much larger than that of the
original film \cite{Fleisch1,Fleisch2,Fleisch3}. Within these crystals
the chance for multiple inelastic scattering events of
$\beta$-particles is enhanced, shifting their energy loss spectrum
towards higher losses. Undiscovered, this shift is faking a lower endpoint 
in the fit which in turn drives $m^2(\nu_e)$through the correlation 
(\ref{gln_corr}) into the unphysical negative sector.

This effect is the stronger, the more the data interval extends towards lower 
energies where it takes in more of
these multiple scattered particles. This trend was clearly seen in our
first publication, already, and attributed to a yet unidentified
additional energy loss component at 75 eV \cite{Weinheim93}. Actually,
this number makes sense to the multiple scattering explanation, since
the average energy loss per scattering event is (34.4$\pm$3.0)~eV
\cite{Aseev} and double scattering prevails in these tiny crystals.
Duly later, however, we learned about the dewetting possibility of
hydrogen films \cite{Leiderer} which was not expected to occur below
the triple point. We were thus forced to study this phenomenon also
for tritium films, determined the decisive activation energy for
surface migration to be 45~K and concluded from that on a dewetting time
constant $\tau_d\gg 1$ year at $T<2$~K \cite{Fleisch3}. The substrate 
temperature throughout running was (1.86$\pm$0.01)~K~\footnote{This number 
corresponds to the reading at the cryostat itself, the absolut precision of 
the temperature is known to 0.1~K.}. Moreover, the source 
section was upgraded to house a larger (2~cm$^2$
instead of 1~cm$^2$) and thicker source ($\approx$ 140 instead of 30~monolayers
of T$_2$) in order to cope with the strong gaseous T$_2$ source of the
competing experiment at Troitsk \cite{Belesev}.  As substrate we have
used again highly oriented pyrolytic graphite (HOPG) which combines
three advantages: \\
(i) ~~low backscattering due to the low Z, \\
(ii) ~atomic flat surface over wide terraces, \\
(iii) high purity \cite{Kettig}. \\
The substrate was glued to the copperhead of the cryostat with the
silverloaded, heat conducting glue H20E (supplied by Polytec, D-76337
Waldbrunn). It withstood cryo as well as baking temperatures of
410~K. 

\begin{figure}
\centerline{ \epsfig{file=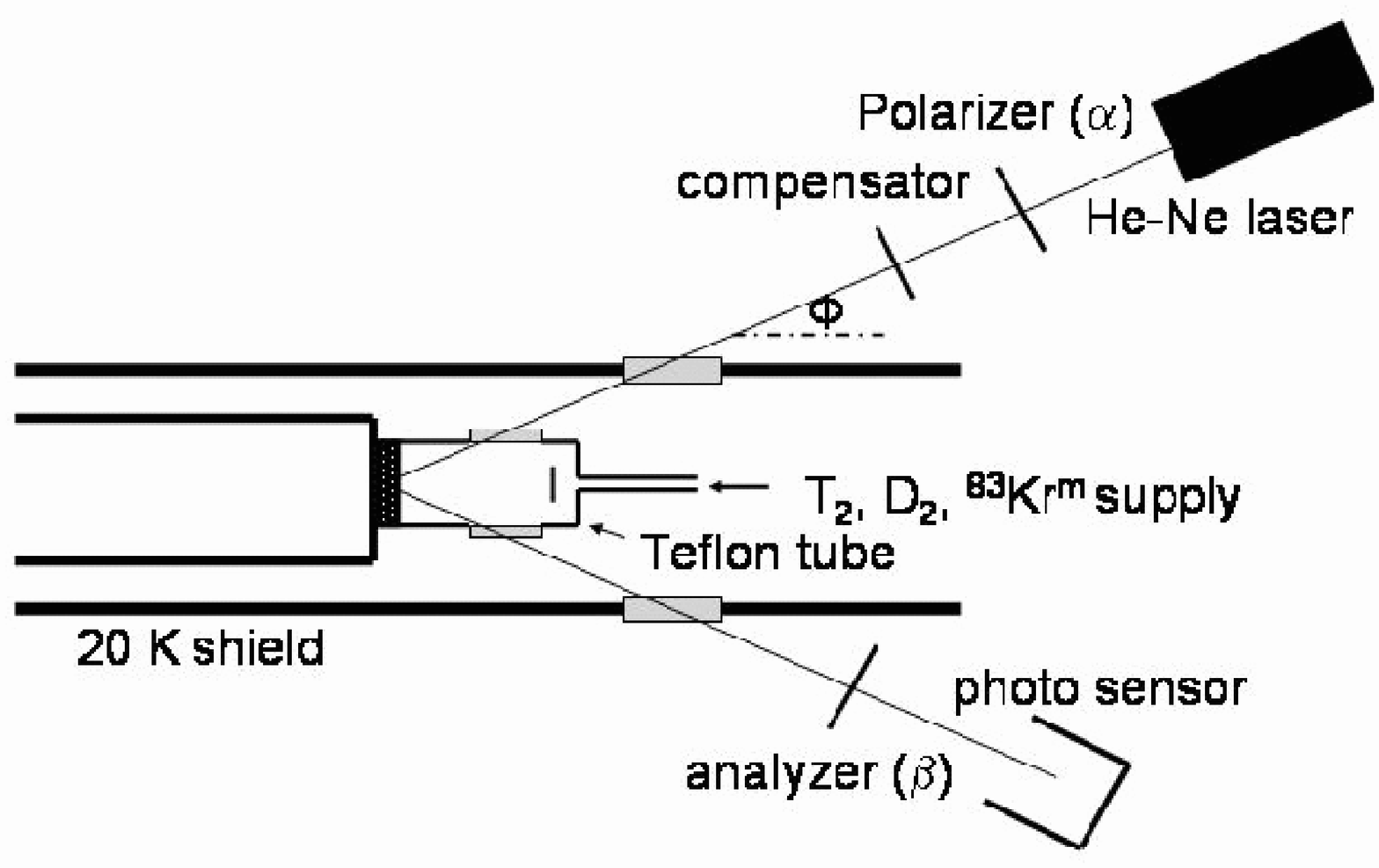, width=0.35\textwidth, angle=-90}}
\caption{Scheme of the tritium source with setup for growing the T$_2$ film 
and controlling its thickness by ellipsometry}
\label{fig:figure4}
\end{figure}

Also the source preparation section has been modified: T$_2$ gas was
released from a heated titanium pellet and fed through a vacuum baked
stainless steel capillary and by help of a mechanical UHV manipulator
into a cold (20~K) radiation shielding tube which surrounded the
substrate (Fig.~\ref{fig:figure4}). The precooled gas then entered a
teflon cup with an inner cross section of 2~cm$^2$ which was pressed
against the substrate. A kind of diffusor at the inlet ensured a
homogenous molecular flow onto the substrate. The isotopic composition
was checked by a quadrupole mass filter.  The isotopic T content of the
individual sources varied between 63\% and 84\%. It was considerably
improved as compared to phase~I.

Radiation shield and evaporation cup were provided with quartz windows
passed by a He-Ne laser beam which monitored on line the growth of the
source film by ellipsometry. For a given polarization status of them 
incident beam, the status of the outgoing beam depends through
Fresnel's formulas on the reflection from both sides of the film and
on the interference of the partial waves. Hence the complex refractive
indices $(n_{\rm f(s)}+ik_{\rm f(s)})$ of the film $(\rm f)$ and the substrate
$(\rm s)$ enter as well as the film thickness $d$ and
the angle of incidence $\phi$. Ellipsometry is performed by help of a
polarizer and a $\lambda/4$ compensator in the incident beam
and an analyzer in the outgoing beam ahead of a photocell 
(Fig.~\ref{fig:figure4}). Fixing the easy axis of the compensator to
$\gamma$ = 45$^\circ$ with respect to the plane of incidence one
searches for that pair of polarizer angle $(\alpha)$ and analyzer
angle $(\beta)$ for which the reflected beam is extinguished. They are
functions of the above parameter set \cite{Azzam}.

\begin{figure}
\centerline{\resizebox{0.45\textwidth}{!}{\includegraphics{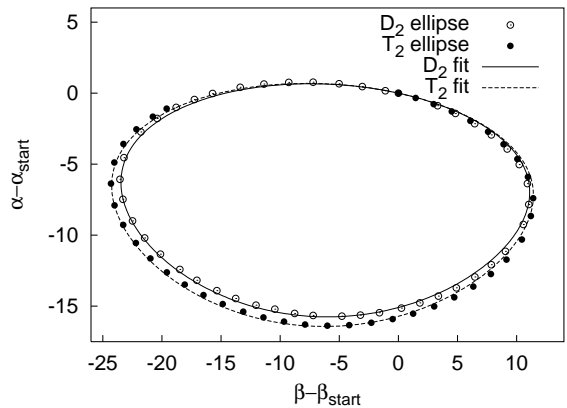} }}
\caption{Control of film growth by ellipsometry for D$_2$ (open circles) and 
T$_2$ (full circles). On the axes are given the corresponding shifts of light 
extinguishing ($\alpha$, $\beta$)- pairs. The lines are fits to the data. 
The loop closes at the first interference order.}
\label{fig:figure5}
\end{figure}

Except for $k_{\rm f} \approx 0$ the indices are not known a priori with
sufficient accuracy. Hence we have grown stepwise rather
thick films of D$_2$ and lately also T$_2$ up to the first
interference order at $d$=4200~\AA, determined $(\alpha,\beta)$ pairs
of extinction (Fig.~\ref{fig:figure5}) and fitted the parameter set to
the data with the results (for the example of Fig.~\ref{fig:figure5})
\cite{BornL}: \\
D$_2$ film: $n_{\rm f}({\rm D}_2) = 1.148,~n_{\rm s}=2.75,~k_{\rm s}=1.34$;
T$_2$ film ($(65 \pm 10 )$\%T, $(35 \pm 10)$\%H): $n_{\rm f}({\rm T}_2) = 
1.156$. $\phi$ is measured geometrically to be (59$\pm 0.3)^\circ$; the fit
yields about equal value and precision for $\phi$.

The T$_2$ films used in the runs are less than 500~\AA~thick and the
measured $(\alpha, \beta)$ pairs cover only a small section of the
full loop, indistinguishable there between D$_2$ and T$_2$.
The only parameter safely extracted from this short section is the
optical film thickness $n_{\rm f} d$. For our experiment, however,
counts the number column density $\rho_N d$, connected to the
refractive index by the Clausius Masotti relation
\begin{equation}
\rho_N=\frac{n^2-1}{n^2+2}\;\,\frac{3}{4\pi\alpha}
\label{gln_elli}
\end{equation}
with $\alpha=0.81(1)$\AA$^3$ \cite{Kolos67} being the polarizability
of hydrogen molecules (for any isotope). From our ellipsometric $n_{\rm f}$
values we thus determine a molar volume of shock-condensed D$_2$:
${V}_{\rm{mol}\, {\rm D}_2, shock}=21.32$ cm$^3$ and of \\ 
T$_2$: $V_{\rm{mol}\, {\rm T}_2,shock}=20.27$ cm$^3$ \cite{BornL}. The
respective literature values for solid (closely packed) D$_2$ and
T$_2$ are 19.95~cm$^3$ \cite{Silvera} and 18.9~cm$^3$ \cite{Fischer}
respectively.  The latter is based on calculations. Hence our
shock-condensed films exhibit a porosity of
\begin{equation}
p=(1-\frac{{V}_{\rm{mol\, c,p}}}{{V}_{\rm{mol,\,
shock}}})=6.4\%~ ({\rm D}_2) ~\rm{and}~ 6.8\%~ ({\rm T}_2).
\label{shock}
\end{equation}
This is considered in our later discussion on energy loss.

Up to the year 2000, we have analysed the ellipsometry of T$_2$ films
using refractive index and molar volume of our shock condensed D$_2$
films. This lead to a systematic underestimation of the T$_2$ column
density by 1\%.  Note, however, that the isotopic composition of our
T$_2$ film varies up to 20\% leaving still a slight residual
uncertainty about the molar volume of the actual film \cite{BornL}.
Moreover, we mention that an alternative calculation of the
T$_2$-column density from the measured count rate leads to qualitative
agreement but is not sufficiently precise for a quantitative check.

The optical quality of the graphite surface apparently deteriorated
somewhat in time, and led to an increase of the relative error of the
film thickness from $\pm$3\% (usual case) to +7\%/-6\% for the
worst case of run Q8 (for more details see \cite{BornB,BornL,Kraus}). 
It is a major contribution to the systematic error through the resulting 
uncertainty in the energy loss within the source.

After film preparation the source is pushed into the front end of a
LHe-cooled chikane, which spans throughout the kinked solenoids S1 und S2. 
It is the second important improvement of the source section. Evaporating 
T$_2$ is adsorbed on its carbon coating and the straight flight into the 
spectrometer is prevented by its 20$^\circ$ bent, whereas the magnetized 
charged particles follow the equally bent field lines. The cryo\-trap 
totally rejected any source dependent background which earlier made up 
half of the background rate, even for much thinner sources.

The cryotrap also suppressed condensation of rest gas (predominantly
H$_2$) from the spectrometer onto the T$_2$ film.  Still we have
observed by an ellipsometric check at the end of a run a certain
growth of the film thickness by 0.14 monolayer/day.  The source
activity, on the other hand, decayed with an apparent lifetime of
about one year. Obviously the recoiling daughter molecules each
sputtered a handful of neighbouring molecules off the source.

\subsection{Electrode and HV system} 
The design of our spectrometer has been aiming at a short, economic
instrument with high resolution, that is a high field ratio 
$B_{\rm max}/B_{\rm a}$. Consequently it has
sharp $B$ gradients which endanger adiabatic motion. Therefore, we
have tried to compensate this drawback by decelerating and
reaccelerating the particles partly in the high field within the
solenoids S3 and S4. This was provided by a series of ring electrodes.

However, field emission by the strong electric field and
particle storage by the strong magnetic field, together favour the
development of plasmas even under UHV conditions. Such plasmas
lead to an untolerable background rate at field settings $B_{\rm
  max}\geq 2$~T. Hence the spectrometer could not be operated up to
its limit of $B_{\rm max}=8.6$~T where adiabaticity would be observed
best \cite{Picard}.

Therefore, the improvement phase also included a redesign of electrodes 
No. E6 to E11 in the high $B$ field \cite{BornB}.
Their number was increased by 2 (E12,E13) in order to
smoothen the potential drops and titanium was chosen instead of copper
for reasons of lower field- and X-ray emission. The latter was
suspected to produce through secondary reactions a background
component observed about 5~keV above the filter potential whose tail still 
extends into the accepted energy window of the detector
\cite{Barth,Goldmann,Picard}. Moreover, these electrodes
were reshaped such that they include about the same magnetic flux
($\approx 6$~Tcm$^2$) everywhere. The new electrode system performed
better in so far, as the mentioned background component disappeared,
indeed, and a breakdown voltage of -30~keV could be reached safely
within a shorter conditioning phase.

But still the plasma induced background rate rose beyond
$B_{\rm max}\approx 2$~T. In the presently running phase~III of the
experiment, which comprises an extensive background exploration and
reduction program to prepare the follow-up experiment KATRIN
\cite{Katrin}, we have removed the multi-electrode system from the high
field region and retained only a few central electrodes, all set to
the full analyzing potential. Now the spectrometer is stable up to the
highest $B$ field, in accordance with earlier experience at the Troitsk 
spectrometer.
For this latter, non-bakable instrument no stable running mode was found
at all with the original multielectrode system \cite{lobashev}.

The filter potential $U_{\rm a}$ was provided by a highly stabilized HV
power supply (model HNC5 30000-5 by Knuerr Heinzinger, D-83026
Rosenheim) directly connected to the central electrode. The potentials
of the other electrodes (requiring less precision) were derived from
$U_{\rm a}$ by a home-made resistive voltage divider. $U_{\rm a}$ was 
monitored and read out continuously by two different systems: 
The first system comprises a high-precision digital voltmeter (model 
DMM 6048 by PREMA, D-55129 Mainz) which was connected to $U_{\rm a}$ via
a precision voltage divider 1:5000 (model KV 50 by Julie Research, New York, 
USA). 
In the second system the voltage $U_{\rm a}$ was divided by 1:50 
by a second voltage divider (model KV 50 by Julie Research, New York, USA)
and the difference to a voltage standard (model 335A by Fluke) 
was measured by a precision digital voltmeter (model DMM 5040
by PREMA, D-55129 Mainz).
The observed short- and long-time fluctuations comply with the specifications 
of the instruments.
To check the HV equipment the K(32) conversion line of $^{83m}$Kr was measured 
before or after each tritium run. The values show a small drift from Q2 to 
Q12 but the difference to the fit values for $U_0$ (given in Table 1) of 
1998-2001 can be summerized as 749.5 $\pm$ 0.5 eV and appear reasonable 
compared to the specification of the HV chan. To control the stability during 
a measurement periode, we analysed shorter time intervals and compared the 
resulting retarding voltages which are found by the fit for the endpoint 
values. They agree within their statistical uncertainties. 
                                                         
The scanning potential in the range -320~V$\leq U_{\rm s}\leq$-20~V
was provided by a fast computer controlled power supply 
model HNC10 3500-10 by Knuerr Heinzinger, D-83026
Rosenheim) and applied to the electrically insulated source. 
A high-precision divider (Fluke) and a high precision digital voltmeter
(model DMM 5017 by Prema, D-55129 Mainz). The minimum negative bias
of -20 V prevents, that recoil ions emmitted from the source are accelerated 
into the spectrometer where they cause a high background rate through rest 
gas ionization \cite{Weinheim93}.

\subsection{Spectrometer vacuum and conditioning} The improvement
program also comprised electropolishing of the spectrometer tank and
its electrodes, in
order to reduce outgassing and field emission but also for removing
any tritium contamination from phase I when we were running without
the protection by the cryotrap. Also the 80~m of getter strip (type ST
707/CTAM/30D by SAES, Milano, Italy), mounted onto the inner surface of
electrode E2 were renewed. It represents a pumping speed of
18~m$^3$/s for hydrogen. The spectrometer was pumped in addition by 2
turbomolecular pumps at 500~l/s each.

Once a year the spectrometer was baked for about a week reaching a
maximum temperature of 330$^\circ$C to 420$^\circ$C for about 24 hours,
at which also the getter was activated. Thereafter the rest gas
pressure (mainly H$_2$) reached a level of better than 10$^{-10}$
mbar.  Although we could not observe any deterioration of the vacuum
in between, the performance of the spectrometer apparently improved after
a very intense rebaking in 2001 in the sense that tiny anomalies appearing in
the spectra of runs Q9 and Q10 in 2000, did not occur anymore, thereafter.

In addition to electropolishing and baking, conditioning of the
spectrometer up to $\pm$30~kV, well above the operating voltage of
18690~V (neg.), proved to be necessary to prevent any sparkings or
minisparkings during runs. The latter are not observed in the electric
circuit but manifest themselves by an outburst of background events
which die out quite slowly such that a whole scan (passing all measurement 
points twice) has to be rejected. In
the runs of 2001 (Q11,Q12), not a single background burst has been
observed. In all earlier runs they appeared about once a week.

\subsection{$\boldsymbol \beta$-detection and data acquisition} 
Transmitted $\beta$-particles are detected by a silicon detector which is
segmented into 5 circular rings of 1~cm$^2$ area each. Usually only the 
three inner segments are considered for data evaluation, the fourth displays 
already enhanced background which increases towards the spectrometer walls. 
The radial segmentation is also useful for accounting for the potential drop 
which occurs in the centre of the analyzing plane and achieves 
4$\cdot$10$^{-5}$ $U_{\rm a}$ on axis.

For phase~II the possibly contaminated old detector was replaced by a
new one with a still thinner dead layer but otherwise identical 
specifications (B1256 by Eurisys Mesures, France).The energy loss in the dead
layer was determined to be about 200~eV at $E$=18~keV, corresponding  to a 
mass layer of about 15~${\rm \mu}$g/cm$^2$. 
The detector together with the attached preamplifiers were cooled down to 
$\approx -80^{\circ}$C. The preamplified signals were fed out of the vacuum  
for further amplification and pulse height analysis. Details are given in 
\cite{Barth97,Wein92}.

\begin{figure}
\centerline{\resizebox{0.45\textwidth}{!}{\includegraphics{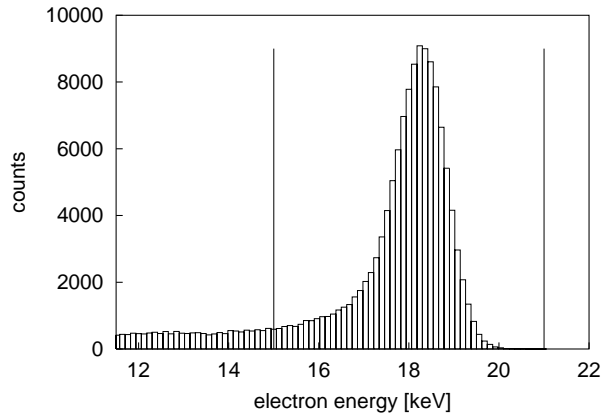} }}
\caption{Detector response to the last 200~eV of tritium $\beta$ decay. The 
perpendicular lines indicate the accepted window (15-21~keV).}
\label{fig:figure6}
\end{figure}

Backscattered electrons are remirrored onto the detector within its 
intrinsic time resolution. They contribute to the low energy tail of the
signal with an energy loss by multiple passage through the dead layer. 
Fig.~\ref{fig:figure6} shows the detector response to the last 200~eV of the 
tritium $\beta$ spectrum. The FWHM is 1.4~keV, the accepted energy 
window 15~keV to 21~keV.
Reinforcement of low level lead shielding and removal of some potassium 
containing material reduced the background from environmental 
radiation by a factor of 3 down to a rate of 4.6$\cdot 10^{-5}$/(s~keV) on 
each segment. Moreover, the vacuum conditions of the detector housing were 
improved to UHV standards in order to allow removal of the thin foil which 
earlier had separated it from the much better spectrometer vacuum. It had 
deteriorated the energy distribution of the passing $\beta$'s \cite{Picard}. 
Instead an open, getter coated tube was installed serving as an active
differential pumping section \cite{BornB}.

\begin{figure}
\epsfig{file=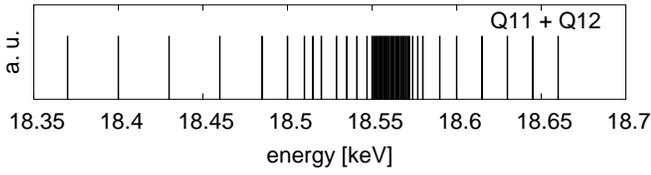, width=0.14\textwidth, angle=-90} 
\caption{Distribution of the 45 measurement points in runs Q11 and Q12. In 
the dense region their spacing is 1~eV.}
\label{fig:figure7}
\end{figure}

At fixed energy window of the detector its efficiency slightly increases 
with $\beta$ energy due to the asymmetric signal slope. By offsetting the 
$\beta$ spectrum at the source the respective coefficient was determined to be
$\alpha_{\rm d}=(4\pm2)\%/\textrm{keV}\;.$ It is considered in the analysis 
withn marginal effect. The low count rate of less than 250~Hz allows to 
acquire the data event by event without suffering substantial losses by signal 
processing and read out, requiring altogether 63~${\rm \mu}$s~\footnote{From 
2000 on this number was decreased to 50~${\rm \mu}$s due to a faster 
computer for data aquisition.}. Pile up 
rejection rises no problem in view of the time constant of 3~${\rm \mu}$s of 
the analogue circuit. Since any enhanced pile up rate points to electronic 
noise or some other pertubation it is recorded in order to reject such 
periods in the off line analysis. The event protocol comprises height and 
real time of the event. Moreover, its time difference to the foregoing event 
is recorded with a resolution of 100~${\rm \mu}$s for the purpose of 
correlation studies. Scanning of the spectrum via the source potential 
$U_{\rm s}$ is PC controlled. 
Usually a measuring time of 20~s per data point was chosen.Their distribution 
has been adjusted to ensure a properly weighted sensitivity to the decisive 
fitting parameters. Fig.~\ref{fig:figure7} shows the example of runs Q11 and 
Q12. Other runs had somewhat different distribution and number of measuring 
points. The potential differences between data points are ramped with
soft slopes over 3~s in order to prevent particle trapping by sharply
rising potential walls. A total scan comprises an up and a down scan.
At the end of each data point the filter and source potentials $U_{\rm a}$,
$U_{\rm s}$ are read out and stored. Moreover a number of other important
control parameters such as the source temperature, the He throughput
through the cryostats, the status of the vacuum system etc. are monitored. 

Any considerable deviation from normal status activates an automatic control 
and safety system which communicates the malfunction as short
message via mobile phone to the operators in charge. It also performs
a safety shutdown if necessary. Vice versa the operators could access
the control system at any time and read out the essential parameters
remotely. Except for serving hours, therefore, the experiment was
running around the clock in a stand alone mode, a necessity in view of
the small crew involved.

\section{Measurements in phase II, 1997-2001}

\subsection{Spectrometer background}

\begin{figure}
\centerline{
\epsfig{file=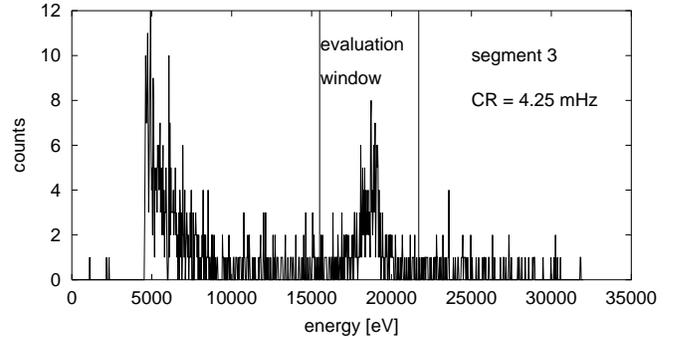, width=0.24\textwidth, angle=-90}
}
\caption{Spectrometer background spectrum, collected for 13~h on detector 
segment 3. The perpendicular lines indicate the accepted energy window for 
evaluation.}
\label{fig:figure8}
\end{figure}

Fig.~\ref{fig:figure8} shows a typical background spectrum from the
spectrometer as measured by the detector with the T$_2$ source closed
off, either mechanically by a valve in the beam line or just with
respect to the $\beta$ particles by a filter potential above $E_0$.
The installation of the bent cryotrap has totally suppressed any source 
dependent background as said before. On top of a smooth continuum one
observes a single peak. A high statistics analysis has shown that its
mean coincides with the filter potential within an uncertainty of about
30~eV. Most probably this peak is a sharp line,
actually, stemming from electrons, produced at low energy somewhere in
the large analyzing volume of the spectrometer or at the surface of
the respective electrodes, and then accelerated by the filter
potential towards the detector or the source. According to (11) this
is possible, if their transverse energy $E_{\perp {\rm a}}$ is less than the
filter width of about 4~eV. Otherwise they will be trapped
magnetically within the $B$ field minimum in the centre.

The rate of this background, which we cannot distinguish from the
$\beta$ particles, ranges from about 12~mHz in the very best cases of
the last runs Q11 and Q12 up to the order of 50~mHz at poorer
performance. The numbers from the similar Troitsk experiment are quite
comparable.

The phenomenology of our spectrometer background has been studied
extensively in a number of thesises over the years
\cite{Picard,Goldmann,Ullrich,Schall,Mueller,Schwamm,Thuem02,Sanchez}.  
The qualitative insight obtained there\-by was very instrumental for
improving to the present satisfactory status. But the various
mechanisms at work are complex, apparently, such that they could not
be identified and disentangled clearly and pinned down quantitatively,
neither by experiment nor by simulation. Only the radical hardware
measures during the presently running, background dedicated phase III
are giving experimental access to a somewhat better understanding of the
underlying mechanisms \cite{Mueller,Schwamm,Thuem02,Sanchez}. Thus routes 
for further background suppression are opened for KATRIN. Since this will be 
subject of a forthcoming paper \cite{flatt04}, we will confine the background 
discussion in this paper to the context of phase~II runs and results. 

From background studies with external $\gamma$ and X-ray sources and from 
coincidence with passing cosmic muons, it seems to be
clear that an important background component -- if not all of the
observed "hard core" of 12~mHz -- consists of secondary electrons
emitted from the inner surface of the large central electrodes. In
perfect adiabatic motion they would spiral along peripheral flux lines
which pass by the detector. However, the actual electromagnetic
configuration with its rather weak central $B$ field in combination with 
radial $E$ field inhomogenities seems to give them a chance to drift into 
the sensitive flux tube on a non adiabatic path. From muon coincidences we 
learned that at least part of them arrive within a few ${\rm \mu}$s.
At UHV conditions, these events cannot be affected any more by rest gas 
collisions. Recently we have found that such electrons can be rejected by a 
grid at some repelling potential \cite{Mueller}. 

Contrary, a single high energy electron, as e.\,g. from T$_2$ decay
within the sensitive flux tube of the spectrometer, may well be stored
magnetically for minutes and cause background events by rest gas
ionisation with an average rate of the order of the observed one at a
vacuum of 10$^{-10}$ mbar. At 10 times higher pressure this background
source can be recognized (and also eliminated at low signal rate) as a
relatively fast chain of correlated events \cite{Belesev}. Also  
mini\-sparks or field emission may end up eventually in such trapped 
high energy electrons.

Guided by such hypotheses we have applied rf pulses of (1.0-1.8)~MHz and
up to 180~V amplitude onto particular electrodes in order to heat up
such trapped electrons stochastically and expel them 
\cite{Weinheim99,Ullrich}. This attempt turned out successful, in fact, when 
the rf pulse was applied to electrode E8 on the detectorside which is at 
a DC potential of 0.86~$U_{\rm a}$. The rf was applied for (1-2)~s each time, 
during the pause when the scanning voltage was changed. In run Q5 for example 
we thus managed to reduce the background rate from an unsatisfactory level 
of 50~mHz down to $\approx$20~mHz \cite{Weinheim99}.

\begin{figure}
 \epsfig{file=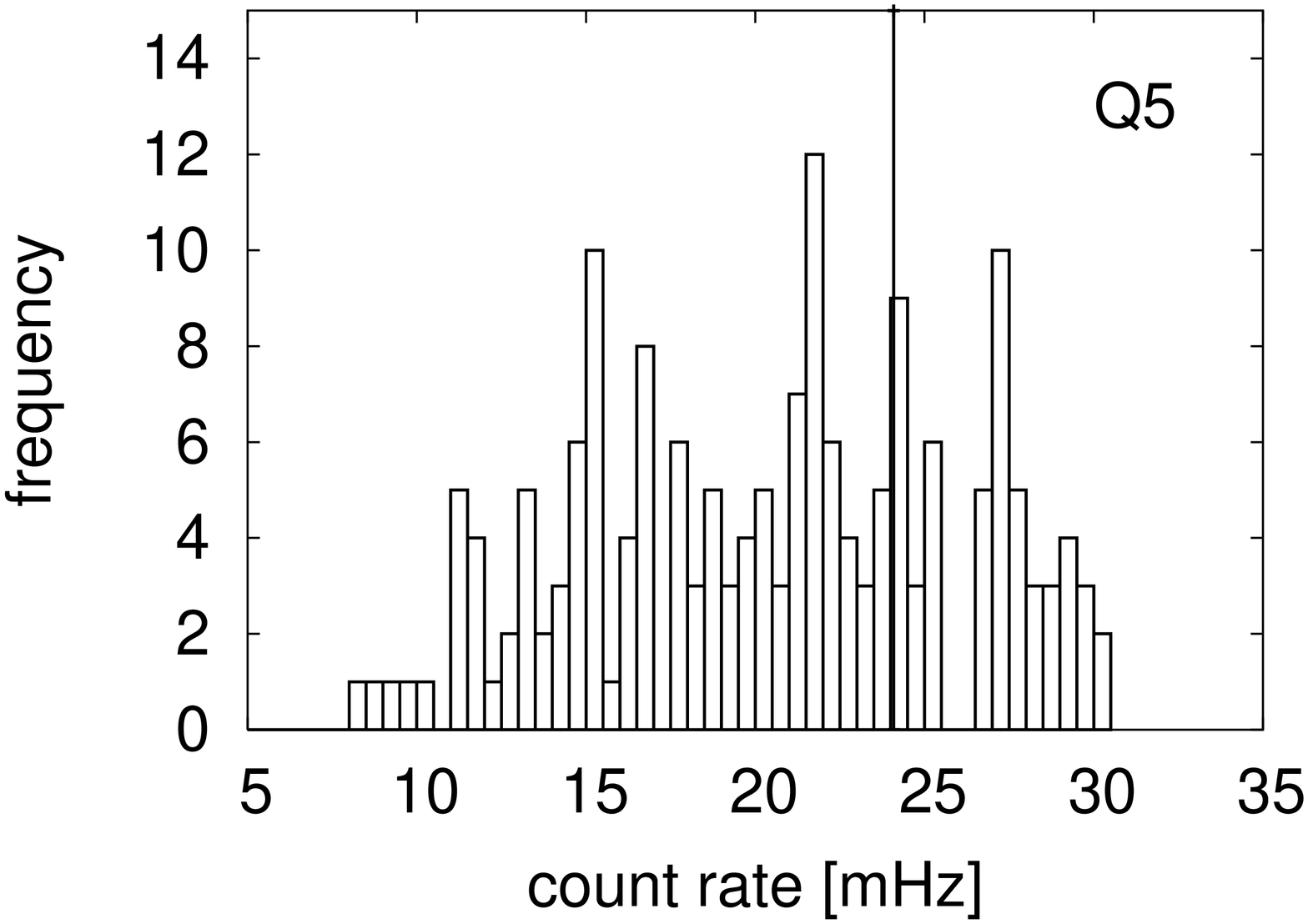, width=0.16\textwidth, angle=-90}
 \epsfig{file=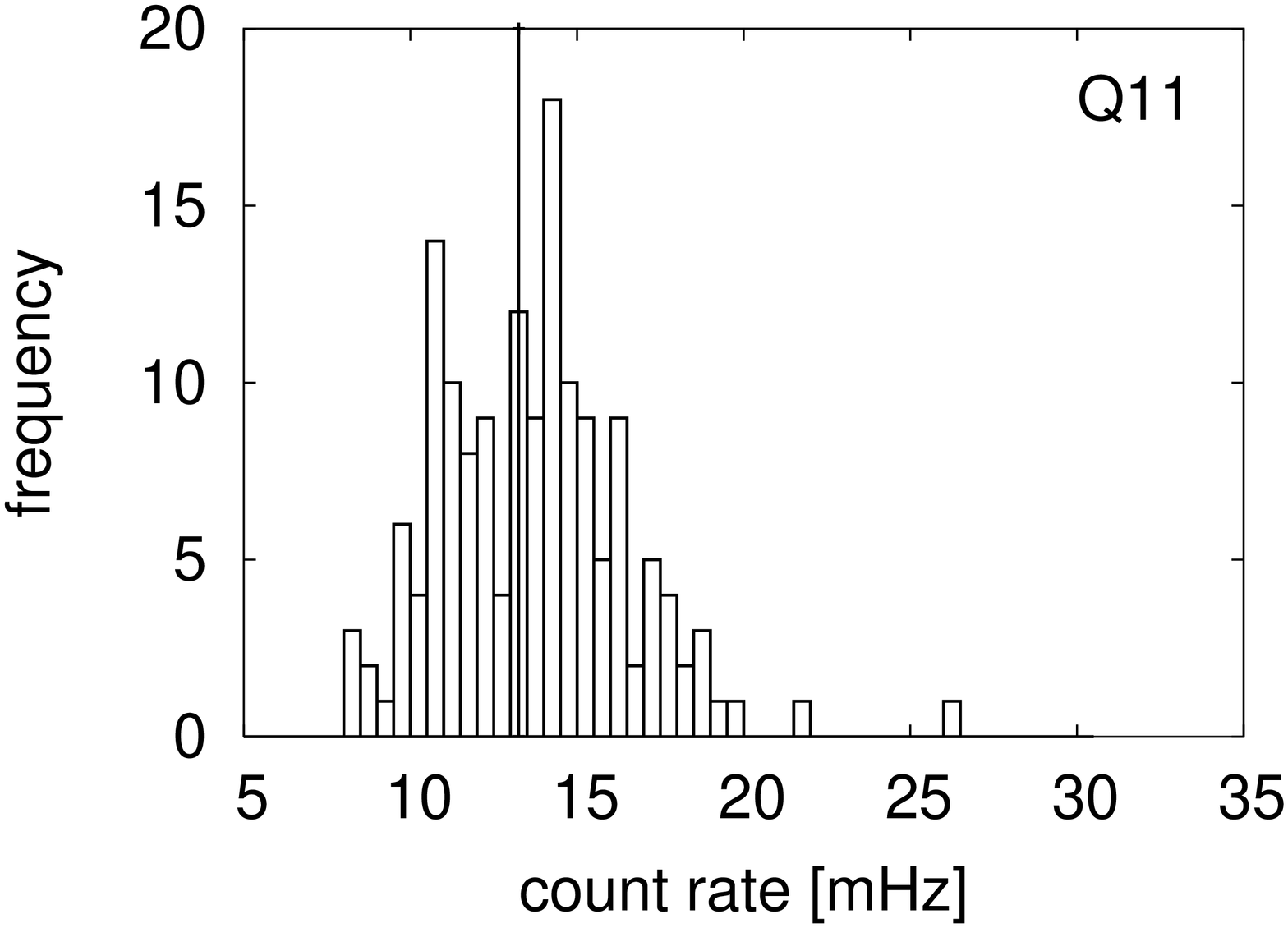, width=0.16\textwidth, angle=-90}
\caption{Histograms of background rates during the measurements Q5 and Q11. 
Each entry is an average over 8 scans, where the measurement time in the 
background region adds up to 600~s for each scan.} 
\label{fig:figure9}
\end{figure} 

Fig.~\ref{fig:figure9} shows two histograms of background fluctuations,
the one from run Q5 with pulsing, the other from run Q11 without
pulsing. In the former case pulsing did reduce an enhanced background
to an acceptable average rate of $\bar{b}=21.6$~mHz.  But the
fluctuation is apparently wider than $\sqrt{bt}$, expected from ordinary 
statistics. This makes sense, since enhanced background rates during the 
occasional presence of ionizing trapped electrons obey a kind of Levi 
statistics with irregular fluctuations. On the other hand, the low and 
steady "hard core" background of $\bar{b}=12.6$~mHz, achieved in Q11, 
displays a nearly ordinary fluctuation.

At UHV conditions, also the build-up and decay of trapped plasmas 
will occur on longer time scales \cite{Picard}. A certain
phase space of electrons can be trapped everywhere within the
sensitive flux tube of the spectrometer, either by magnetic mirroring
at both necks of the magnetic bottle or only at one of them and
electrostatic reflection at the central filter potential.

Outside the sensitive flux the stepwise increase of the
diameter of electrodes forms small equipotential corners crossed twice
by magnetic field lines. This is the electromagnetic configuration of
Penning traps. Electrons released from plasmas in such traps may eventually 
find their way to the detector. Also ions from such plasmas, positive as
well as negative ones, can contribute to the backgroud by secondary
reactions as said before.

If trapped plasmas are fed at least partly by $\beta$ particles
from the source, then the background they produce will depend on the
$\beta$-flux therein, which varies with the scanning potential of the
source. Such kind of cross talk between scanning and background could
be the origin of the tiny residual spectral anomalies which have been
observed in few of the runs and reported already in \cite{Weinheim99}.
Their amplitude is on the level of mHz, i.e. of order 10\% of the
total background rate. The slow build-up and decay rates of stored
particles may also give rise to the hysteresis of these anomalies
which has been observed between up and down scanning in run Q9.

In a pulsed mode of running, the technique of which has been described in
\cite{bonn99}, we have searched for a background dependence on the
scanning voltage in a time window ranging from a few ${\rm \mu}$s to 30~min
after shutting off the signal rate by a positive pulse on the source.
Within statistical limits of 1~mHz we did not observe any correlation
between the background rate in the pauses and the signal rates in the open
phases of the spectrometer \cite{Schall}. This negative result does
not really contradict our presumptions on the possible origin of the 
residual spectral anomalies, since these have only rarely occured at non 
optimal conditions.

Total stability of the background rate over the full running period
has been observed for runs Q11 and Q12.

\subsection{Discussion of Runs Q1 to Q12}
In Table 1 we have listed characteristic parameters of all 12 runs Q1
to Q12 performed with the improved setup in the period 1997 to 2001.
They covered an interval from 18370~V up to 18860~V, i.e.
from 200~eV below to 90~eV above $E_0$ (Fig.~\ref{fig:figure7}).

\begin{table}[h]
\center
\caption{Parameters for measurements Q1-Q12 of phase II. 
$\Theta_{\rm max}$ = maximal opening angle, $t$ = running time, $pt$ = number 
of different measurement points, $ft$ = film thickness, 
$\bar{b}$ = average background rate, $m^2(\nu_e)$ = fit result for the 
last 70~eV and $U_0$ = retarding voltage at which fit finds endpoint value.} 
\label{table}
\center
\hspace{-0.8cm}
\begin{tabular}{|p{0.4cm}|p{0.45cm}|p{0.2cm}|p{0.2cm}|p{0.5cm}|p{1.1cm}|
p{1.3cm}|p{1.9cm}|} \hline
{$no.$} & {$\Theta_{\rm max}$} & {$t$ [d]} & {$pt$} & $ft$ [nm] & 
{$\bar{b}$ [mHz]} & {$m^2(\nu_e)$ [eV$^2$]} & {$U_0$ [V]} \\ \hline
Q1  & 45$^\circ$ & 6   &   & 20.8 &  & \multicolumn{2}{|c|}{test measurement} \\ \hline
Q2  & 45$^\circ$ & 26 & 50  & 96.7 & 16.7$\pm$0.3 & $-11.2\pm$6.0 & 
18573.5$\pm$ 0.3 \\ \hline
Q3  & 45$^\circ$ & 24 & 64  & 49.3 & 12.7$\pm$0.2 & $-14.8\pm$4.6 & 
18574.0 $\pm$ 0.2 \\ \hline
Q4  & 45$^\circ$ & 38 & 64  & 49.5 & 11.7$\pm$0.2 & $-3.9\pm$4.7  & 
18574.5 $\pm$ 0.2 \\ \hline
Q5  & 45$^\circ$ & 46 & 64  & 47.5 & 21.6$\pm$0.2 & $-3.5\pm$6.0  & 
18574.4 $\pm$ 0.2 \\ \hline
Q6  & 62$^\circ$ & 38 & 33  & 43.0 & 12.5$\pm$0.2 & $+0.4\pm$7.2  & 
18575.7 $\pm$ 0.2 \\ \hline
Q7  & 62$^\circ$ & 29 & 33  & 43.2 & 14.3$\pm$0.2 & $-2.4\pm$4.9  & 
18575.4 $\pm$ 0.2 \\ \hline
Q8  & 62$^\circ$ & 54 & 39  & 45.5 & 16.5$\pm$0.2 & $-0.9\pm$4.8  & 
18576.2 $\pm$ 0.3 \\ \hline
Q9  & 62$^\circ$ & 56 & 39  & 44.4 & 18.6$\pm$0.3 & $-10.9\pm$3.2 & 
18575.1 $\pm$ 0.2 \\ \hline
Q10 & 62$^\circ$ & 35 & 45  & 45.5 & 16.6$\pm$0.3 & $-6.1\pm$4.8  & 
18574.6 $\pm$ 0.2 \\ \hline
Q11 & 45$^\circ$ & 31 & 45  & 48.2 & 12.6$\pm$0.2 & $+1.3\pm$5.8  & 
18576.7 $\pm$ 0.2 \\ \hline
Q12 & 62$^\circ$ & 19 & 45  & 48.5 & 12.6$\pm$0.2 & $-1.0\pm$6.0  & 
18576.6 $\pm$ 0.2 \\ \hline 
\end{tabular}
\end{table}

We should mention that we fitted the parameter $m^2(\nu_e)$ as really free 
parameter without constraints.
In order to account for statistical fluctuations of the data the fitting 
routine\footnote{The shape of the $\beta$ spectrum near the endpoint is 
mainly defined by the factor $(E_0 - E) \sqrt{(Q-E)^2 - m^2(\nu_e)c^4}$
which can be expanded for $(Q-E) \gg m(\nu_e) c^2$ into $(E_0 - E)^2 - 
m^2(\nu_e)/2$. Therefore for a neutrino mass around zero any fluctuation 
of the count rate downwards yields a positive value for $m^2(\nu_e)$ and
vs. any fluctuation upwards should result in a negative value of the 
parameter $m^2(\nu_e)$.} requires a mathematical continuation of the 
spectrum into the region $m^2(\nu_e)<0$ which provides a symmetric $\chi^2$ 
parabola around $m^2(\nu_e)=0$ for a statistical data sample. 
This purpose is fulfilled quite well by introducing factors $f_i$ to 
each electronic final state in (\ref{gln_rate}) defined by
\begin{eqnarray}
f_i=\Theta(-m^2(\nu_e)) \Theta(\epsilon_i+\mu) (1+ \frac{\mu} {\epsilon_i} 
e^{-(1+ \epsilon_i / \mu ) } ) \nonumber \\
+\Theta(m^2(\nu_e)) \Theta(\epsilon_i-m(\nu_e) \textrm{c}^2)
\label{spektrumcorrec}
\end{eqnarray}
with $\epsilon_i=E_0-V_i-E$, $\mu=-0.66$ $ m^2(\nu_e)$c$^4$ \cite{Weinheim93}. 
For negative $m^2(\nu_e)$ they stretch the spectrum smoothly beyond the 
respective endpoint up to $E_0-V_i+\mu$. 
The negative $m^2(\nu_e)$ sector might also be fitted by a physical model, 
namely the $\beta$ spectrum arising from tachyonic neutrinos \cite{cibo}. 
But this point would come up only in case of an unambigous experimental 
negative $m^2(\nu_e)$ result. Summaryzing, negative values for the fit 
parameter $m^2(\nu_e)$ are not necessarily unphysical but should obtained 
within statistical limits as result of an unconstraint fit in 50\% of all 
data sets if the neutrino mass is around zero.

In the following we will report on each of the 12 runs performed in
phase II irrespective of wether it has been selected for the final
data set. Thus we take the chance to discuss on the given example
carefully experimental effects which might lead to spectral anomalies
and systematic uncertainties if undiscovered.

\textbf{Q1 and Q2}: The first run Q1 was devoted to a short test
experiment with a relatively weak T$_2$ source. We observed in
particular that the background due to T$_2$ evaporation into the
spectrometer and decay therein had disappeared as said above.
Encouraged by this success we have produced a very thick source Q2 of
967~\AA~corresponding to 284 monolayers. At this thickness the rate
of $\beta$ particles which leave the source without energy loss is already 
close to its maximum possible value obtained from an infinitely thick source. 
Running this high source activity turned out to be smooth and stable without 
any source dependent background problems.

However, the analysis of the data revealed an average
shift of the endpoint by -3~eV. This effect was then systematically
investigated by freezing $^{83m}$Kr activity on top of T$_2$ films 
and measuring precisely the energy of its 17.8~keV conversion line as
function of the film thickness. This way we have discovered that
the film charges up positively by 21.2~mV per monolayer. The
corresponding electric field strength of 62.6~MV/m is necessary to
release the positive charges, left over from $\beta$ decay, from their
trapping potential within the T$_2$ lattice. These first experiences with 
the improved set up have been communicated in \cite{Barthetal98}. A thorough 
analysis of source charging is given in \cite{auflad}. The linear increase of
the charge up voltage throughout the film has to be folded into the
transmission function and results in a broadening in addition to an
average shift. A systematic uncertainty in the broadening effect of Gaussian
shape with variance ${\rm \sigma}^2$ would yield an uncertainty in 
$m^2(\nu_e)$ by $-2 \cdot {\rm \sigma}^2$ \cite{Wilkerson}. Therefore, we 
have reduced the film thickness by a factor of 2 in later runs. Moreover the 
uncertainty of the energy loss weighs heavier in a thick source than in 
thinner ones. Still, as compared to phase I results, the analysis of Q2 
led to a reduction of the unphysical negative $m^2(\nu_e)$ value by an order 
of magnitude (see Fig.~\ref{figure13} in sec. 5).

\textbf{Q3 and Q4}: One might have expected that the residual small
negative $m^2(\nu_e)$ of order -10 eV$^2/{\rm c}^4$ still observed in run Q2
would disappear with thinner sources.  However, the following runs Q3
and Q4 showed the same problem at similar size but with somewhat
different dependence from the fit interval. Moreover, the clearly
enhanced $\chi^2$ values obtained in fitting Q2, Q3 and in particular
Q4 pointed to some residual spectral anomalies in the data 
(see Fig.~\ref{figure13}). The Troitsk group had already reported on a 
step like anomaly which appeared in their integral spectra with an amplitude 
of order 10$^{-10}$ of the total decay rate and at variable positions in the 
range from 5 to 15~eV below the endpoint. The change in time of the positions 
of these steps seemed to be compatible with a half year period, even
\cite{lobashev99}. If attributed to a general physics phenomenon, e.g. a
monochromatic line in the $\beta$-spectrum of T$_2$, it should appear
in our spectra equally. Fitting such a step into the Q4 spectrum led
to a significant reduction of $\chi^2$, and lifted $m^2(\nu_e)$ to an
acceptable value of (-1.8$\pm$5.1$_{\rm stat} \pm 2.0_{\rm sys})$ eV/c$^2$,
indeed \cite{Weinheim99}. Also the position at 13~eV below $E_0$ and
the amplitude of the step of about 6~mHz accorded to the Troitsk
picture. In the following runs it was an important issue, therefore,
to investigate further and - if possible - eliminate these residual
spectral anomalies.

\textbf{Q5 to Q8}: Ahead of run Q5 we found out that an enhanced and
fluctuating background rate could be reduced essentially by applying
an rf pulse to electrode E8 on the detector side as has been reported
above. In particular Q5 profited from this procedure; it reduced the 
averaged background rate from 50 mHz down to 21.6~mHz. Q5 was also the
first run whose analysis did not reveal any spectral anomaly any more
but yielded an $m^2(\nu_e)$ value compatible with zero at good $\chi^2$
for any data interval (see Fig.\ref{figure13}). 

Still the background rate was higher in Q5 than in the foregoing runs which 
suffered from slight residual anomalies. Apparently these anomalies do not 
correlate necessarily to a higher average background; but a correlation 
between background events and the operating cycle of the spectrometer -- 
which clearly produces an anomaly -- can occur also at low average 
background. At that point we recall that the anomalous count rate does not 
exceed a few mHz and hence constitutes at most a small fraction of the 
background rate. On the other hand the facts seem to corroborate the 
assumption that the removal of trapped electrons by pulsing also may brake 
the correlation between background and operating cycle.
Run Q5 was the basis of our result $m^2(\nu_e)=(-3.7 \pm 5.3_{\rm stat} \pm
2.1_{\rm sys})$ eV$^2$/c$^4$ with the limit $m(\nu_e)<2.8$ eV/c$^2$ 
(95\%~C.L.) published in 1999 \cite{Weinheim99}.
But also the series of data collected in the runs Q3 to Q5 could be
analyzed successfully in a so-called "15~eV analysis". Besides the background 
region above $E_0$, only the last 15~eV of the spectrum were considered here 
together with two more data points further down at 18470~eV and 18500~eV 
respectively; they were necessary to fix $E_0$ with sufficient precision. 
Thus the "troublesome" region of anomalies was excluded mostly from the fit.
The result was $m^2(\nu_e)$ = $(-0.1 \pm 3.8_{\rm stat} \pm 1.8_{\rm sys}) 
$~eV$^2$/c$^4$, which leads to an upper limit of $m(\nu_e)<$ 
2.9~eV/c$^2$ (95\%~C.L.) \cite{Weinheim99}.

In between runs Q5 and Q6 the spectrometer was baked again to a
maximum temperature of 394$^\circ$C and HV conditioned. The procedure
resulted in a background reduction down to 12.5~mHz observed in Q6.
This rate was independent of pulsing as proved by the no pulsing mode which 
ran alternatively every second day. Q7 was running in the same alternating
manner. Without pulsing the background had now increased to 14.7~mHz
(the number given in Table 1), whereas it remained essentially
stable at 12.7~mHz in the pulsing mode. Permanent pulsing was applied to run
Q8, the background had increased further to 16.5~mHz. Note that these
numbers are averages over the full running period. Without pulsing the
background was slowly rising in real time and had to be set back by
reconditioning the electrode system a few times during a run.

On the thirty-first day of run Q8 the "apparent" lifetime of the
source (as measured from the course of its activity) increased from
300 to 620 days. The event was caused by a sudden coverage of the
source with a couple of monolayers of H$_2$, which had been collected before 
from the residual rest gas onto the shield in front of the
source. From there it was released then during a short cooling failure
of the shield and partly recollected onto the still cold source. The
data from the remaining period yielded a significant negative
$m^2(\nu_e)$ value, caused by the additional energy loss in the H$_2$
cover. Also under regular conditions the shield
could not completely prevent a slow and steady condensation of residual H$_2$
onto the source. A daily coverage by 0.3 monolayers was estimated from
ellipsometry (see above) and considered in the analysis causing a
small, still significant effect (see below).

The analysis of the data set Q6 to Q8 yielded stable fit results of
$m^2(\nu_e)$ close to zero at any data interval and with good $\chi^2$
(see Fig.~\ref{figure13}). Together with Q5 it improved the result, to:

\begin{equation} 
m^2(\nu_e)=(-1.6 \pm 2.5_{\rm stat} \pm 2.1_{\rm sys})~{\rm eV}^2/{\rm c}^4,
\label{gln_mainz99}
\end{equation} from which an upper limit of $m(\nu_e)<2.2$ eV/c$^2$ 
(95\%~C.L.) was extracted. This result has been communicated at the Neutrino 
2000 conference \cite{Bonn} and cited thereafter frequently.

\textbf{Q9 to Q10}: The long period of data collection in runs Q5 to
Q8 was followed by a number of systematic background studies
\cite{Ullrich,Schall} the results of which have been summarized above,
already. Thereafter tritium measurements were immediately resumed in
fall 2000 without a break for extended maintenance. Since the last
baking of the spectrometer, which apparently had not reached the
temperature of earlier ones, 6 months had elapsed. Although running
was quite smooth at a moderate background rate of 20~mHz, the on line
analysis of Q9 data showed rather soon a reappearance of slightly
negative $m^2(\nu_e)$ values around -10~eV$^2$/c$^4$. Nevertheless
we continued measuring since parallel runs were foreseen at Troitsk in
order to check, whether any Troitsk anomaly would appear synchronously
in both experiments. We also refrained from any interference by
reconditioning the electrodes like in earlier runs.
Rather we decided to watch how the running conditions and results
would develop in time in the two alternating modes of cleaning and not
cleaning the spectrometer from stored particles by rf pulses.

\begin{figure}
 \epsfig{file=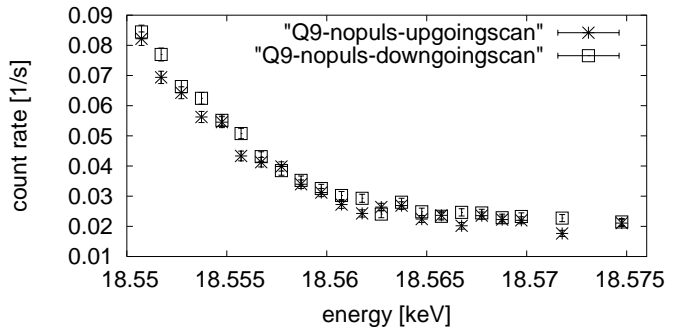, width=0.25\textwidth, angle=-90} 
\caption{Hysteresis effect in run Q9 during periods without background 
suppression by pulsing. Squares correspond to scanning towards the endpoint, 
crosses to the opposite direction.}
\label{fig:figure10}
\end{figure}

An apparent anomaly of Q9 is a hysteresis of countrate between up and
down scanning: It is visible in the raw spectra already
(Fig.~\ref{fig:figure10}). Analysis shows that the effect was much
stronger in the non-pulsed than in the pulsed mode. The effect
also diminished when the final approach to each measuring point was
performed always from the same side, namely from a higher voltage
level; that means, upramping was performed by first overshooting the
proper value and then pulling back as if it were downramping. The
hysteresis clearly indicates the presence and influence of stored
particles, whose accumulation and/or loss conditions correlate by some
mechanism to the setting of the spectrometer. The instability of
background conditions introduced this way is also witnessed by an
enhanced scatter in Q9 as discussed above and shown in 
Fig.~\ref{fig:figure9} for the example of Q5. 
Even if the exact mechanism has not been cleared up yet in detail, the 
knowledge of its phenomenology is already very important for the 
experimentalist and enables to prepare cautious countermeasures.

In the following run Q10 the measuring points have not been addressed
in a monotonous voltage sequence but in a random generated one.
Unfortunately, the software did not allow for a fresh random choice
each scan without major changes. But the same sequence was repeated
forth and back for the whole run. Still this was sufficient to
suppress the conspicious hysteresis of Q9, not so the occurrence of
trapped particles as such, however. Hence also the Q10 spectrum is
slightly impaired by the imperfect running conditions, leading to a
negative $m^2(\nu_e)$ result in summary (see Table 1). Q9 and Q10 data
have also been analyzed in weekly time bins. They show fluctuations of
the $m^2(\nu_e)$ fit result which exceed statistical limits. Hence Q9 and 
Q10 data were not considered in the final fit.

In spite of being slightly impaired, the Q9 and Q10 spectra did not
display any indication of a step like Troitsk anomaly, neither in the
full data sets, nor in binned ones. This holds in particular for the
time bins 6.12.-13.12. and 22.12.-28.12.2000 where the Troitsk experiment
was running in parallel to ours. In both periods Troitsk, however,
observed the sudden outburst of significant steps. In the second period it 
even reached an amplitude of 14~mHz (discussed further below).

\textbf{Q11 and Q12}: The experience of Q9 and Q10 has retaught us the
importance of optimal maintenance ahead of running, although the
vacuum had been fine all the time.  Changes in surface conditions seem
to rule field emission of electrons and/or ions which fill residual traps 
and interfere with the measurement. Hence the spectrometer was rebaked at 
maximum temperature of $373^\circ$C for 13~h. Thereafter, the electrodes 
were conditioned up to $\pm$30~keV with reinforced sparking at a residual 
hydrogen pressure of 10$^{-7}$~mbar, obtained by heating up the SAES getters 
at closed turbo pumps. Also the source section was thoroughly maintained 
including exchange of source substrate and T$_2$-pellet, baking, etc.

The efforts were rewarded with two absolutely clean and quiet runs Q11
and Q12. From altogether 1620 scans 1580 passed all control criteria
in the analysis; among the few rejected runs prevailed incomplete ones
due to some peripheral technical problem or intervention. The
background rate was stable and further reduced by 20\% to 12~mHz
without the necessity of pulsing off stored particles. In Q11 we
applied again a random sequence like in Q10, but returned to monotonous
scanning in Q12.  In the last week of Q12 we ran in a slow scanning
mode at 900~s per point instead of the usual 20~s interval. This way we
searched for possible correlations of rates to scanning steps on an
extended time scale but could not identify any.

\section{Analysis of data}
The way we analyze our data has been described before already
\cite{Weinheim93,Weinheim99,Bonn} and will be shortly resumed. In the 
meantime, we have developed certain refinements, which we also like to apply 
to the already published data resulting in slight changes of the results.

\subsection{Raw data selection}
The raw data of a run consist of a large number of single event
protocols (see above), grouped into single measurements of 20~s at 
particular voltage settings. By help of the CERN routine PAW
the raw data could be visualized in plots performing cuts of data and
correlations of parameters. Outbursts of countrate, e.g.  caused by
some sparking, were identified and rejected manually this way. In fact
the total scan was rejected in case of these rare events since they were
followed by a longer "afterglow" of background events. In the latest
runs they did not occur at all. Also other obvious malfunctions were
identified and rejected this way. This first visual data screening was
followed by an automatic one which identified for each single
measurement significant deviations of the voltage readings from their
nominal value or their average. If they exceeded 0.1~V, the
measurement was rejected. This made sure that even at the highest
signal rate deep in the spectrum the corresponding signal deviation is
less than 0.3${\rm \sigma}$. Moreover the program rejected single
measurements when they contained more than 10 pile up events. For the
remaining data it performed an automatic dead time correction reaching
a level of 1\% for the highest count rates. Altogether, the
percentage of rejected data ranged from 2\% to 6\% for individual
runs.

\subsection{The fit function and the response function}
\label{sec:fitres}
The data from runs Q2 to Q12 were fitted each by a fit function $F(U)$,
which is a convolution of the primary spectrum (2) with the response
function $T'(E,U)$ of the apparatus plus a constant background $b$:
\begin{equation}
F(U)=\int {(R(E)T'(E,U)}\textrm{d}E)+b=R\otimes T'+b.
\label{fitfunction}
\end{equation}
$T'(E,U)$ is again a fivefold convolution of the transmission function 
(\ref{gln_trans}), the energy loss function in the film $f_{\rm loss}$, 
the charge up potential in the film $f_{\rm charge}$, the backscattering 
function from the substrate $f_{\rm back}$, and the energy dependence of the
detector efficiency $f_{\rm det}$:
\begin{equation}
T'(E,U)=T\otimes f_{\rm loss}\otimes f_{\rm charge}\otimes f_{\rm
  back}\otimes f_{\rm det}.
\label{tstrich}
\end{equation}

\begin{figure}
\epsfig{file=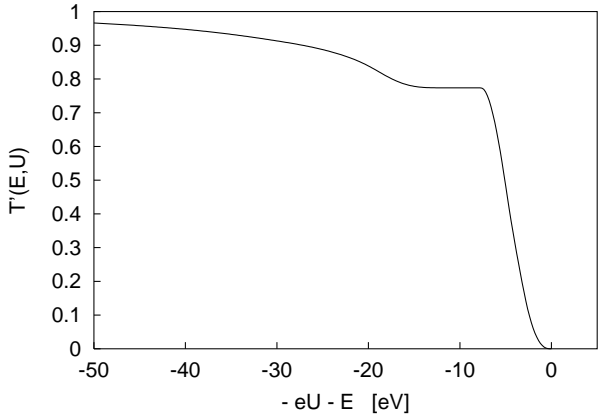, width=0.45\textwidth} 
\caption{Convoluted and normalized response function of the apparatus to 
electrons emitted at energy $E$ and analysed at the filter voltage $U$. The 
filter width ${\rm \Delta} U$ is set to 4.8~V, the maximum starting angle to 
$\Theta_{\rm max}=$ 45$^{\circ}$ and the convolution is calculated for a
source thickness of 490~\AA. }
\label{fig:figure11}
\end{figure}

\begin{figure}
\epsfig{file=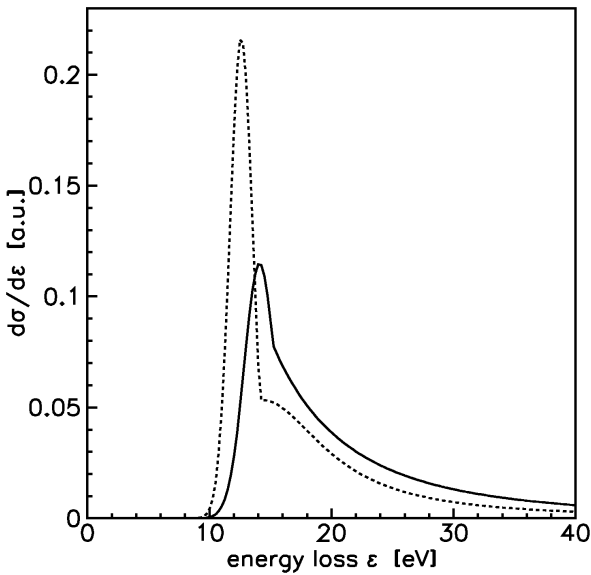, width=0.45\textwidth} 
\caption{Differential inelastic cross sections for 18.5~keV electrons 
scattered from gaseous hydrogen (dashed line) and quench condesend 
D$_2$ (solid line) \cite{Aseev}.}
\label{fig:figure12}
\end{figure}

It is plotted in Fig.~\ref{fig:figure11} for an electron starting somewhere in
the source with energy $E$ and analyzed at a filter setting 
$U$=$U_{\rm a}-U_{\rm s}$. 
Its structure is dominated by $T\otimes f_{\rm loss}$. The filter opens
for the elastic component when the retarding potential $-eU$ balances the
particle energy, reaching a first plateau 8 eV below.  At that point
the elastic component is fully transmitted. The slope of $T'$ is
stretched with respect to that of $T$ by $f_{\rm charge}$ which
spreads over 2.8~eV for a 140 monolayer source. The second, smaller and
softer uprise results from integrating up the inelastic spectrum
$f_{\rm loss}$. At a setting of $-e\cdot (U+E)= -50$ eV electrons are 
transmitted to the detector with a chance of 98\% already. $f_{\rm loss}$ 
has been determined in parallel for gaseous hydrogen as well as for shock
condensed films (actually D$_2$) with the $\beta$ electron spectrometers at
Troitsk and Mainz, respectively \cite{Aseev}. For the condensed case,
the differential cross section d$\sigma$/d$E$ was approximated by two
model functions, a Gaussian peak at an energy loss of 14.1~eV followed
by a third order hyperbola (Fig.~\ref{fig:figure12}). The total inelastic 
cross section was found in this case to be 
$\sigma_{\rm tot}=(2.98\pm0.16)\cdot 10^{-18}$ cm$^2$ with a mean energy 
loss of (34.4$\pm$ 3.0)~eV. As compared to the gaseous phase, it is found that
the excitation peak is shifted upwards by 1.5~eV and the total cross
section lowered by 13\%. The shift is also confirmed by quantum chemical 
calculations \cite{Aseev}. If we define an inelastic scattering coefficient 
$K$ of the film by 
\begin{equation}
K=\sigma_{tot} \cdot \rho_N \cdot \frac{l}{\cos\Theta }
\label{coeffk}
\end{equation}
with $l/\cos \Theta$ being the actual path length of a $\beta$
particle through the film, then the probability of scattering $n$
times is given by a Poisson distribution
\begin{equation}
P_{\rm ne}=e^{-K}K^{n}/n!.
\label{poisson}
\end{equation}
At the given film thickness it is sufficient to take multiple scattering up 
to third order into account. The response function is obtained for each 
tritium film layer at a certain electrical potential - defined by 
$f_{\rm charge}$ --  by an appropriate convolution of Poisson distributions 
(\ref{poisson}) over energy loss and path length \cite{Aseev}. 
Running at different $\Theta_{\rm max}$ changes the response function, thus 
requiring separate evaluation of runs.

Backscattering from the graphite substrate is quite small.
Simulations have shown that its spectrum may be approximated within
the interesting interval of 200 eV below the starting energy
$E$ by a constant pedestal of relative amplitude
\begin{equation}
\alpha_{\rm back}=3.1\cdot 10^{-5}/\textrm{eV}
\label{alphaback}
\end{equation}
with respect to a $\delta$- function at $E'=E$. The latter represents the
transmission probability for forward emission \cite{Fleisch1}. The
number given above applies to $\Theta_{\rm max}=60^\circ$. It
decreases for a narrower transmission cone of $\Theta_{\rm
  max}=50^\circ\; \textrm{to}\; 2.3\cdot 10^{-5}/{\rm eV}$. The simulations
have been checked by test experiments with K-conversion electrons
from $^{83m}$Kr decay \cite{Fleisch1}.

Since the back scattering effect is small it is sufficient to replace the exact
convolution procedure by a simple correction factor
\begin{equation}
f_{\rm back,corr}=1+\alpha_{\rm back}(E+eU).
\label{fbackcorr}
\end{equation}
The second term stands for the integral of the backscattered spectrum
over the width $(E+eU)$ of the transmission window of the spectrometer. Also 
the folding with the energy dependent detector efficiency in
(\ref{tstrich}) can be replaced by applying a simple correction factor
\begin{equation}
f_{\rm det,corr}=1+\alpha_{\rm d}(E+eU)
\label{fdcorr}
\end{equation}
with the coefficent mentioned before
\begin{equation}
\alpha_{\rm d}=(4\pm 2)\cdot 10^{-5}/eV.
\label{alphad}
\end{equation}

The fit is then performed with $m^2(\nu_e), E_0$, $b$ and a signal amplidtude 
as free parameters.

\subsection{Fits of individual runs}
\begin{figure}
 \epsfig{file=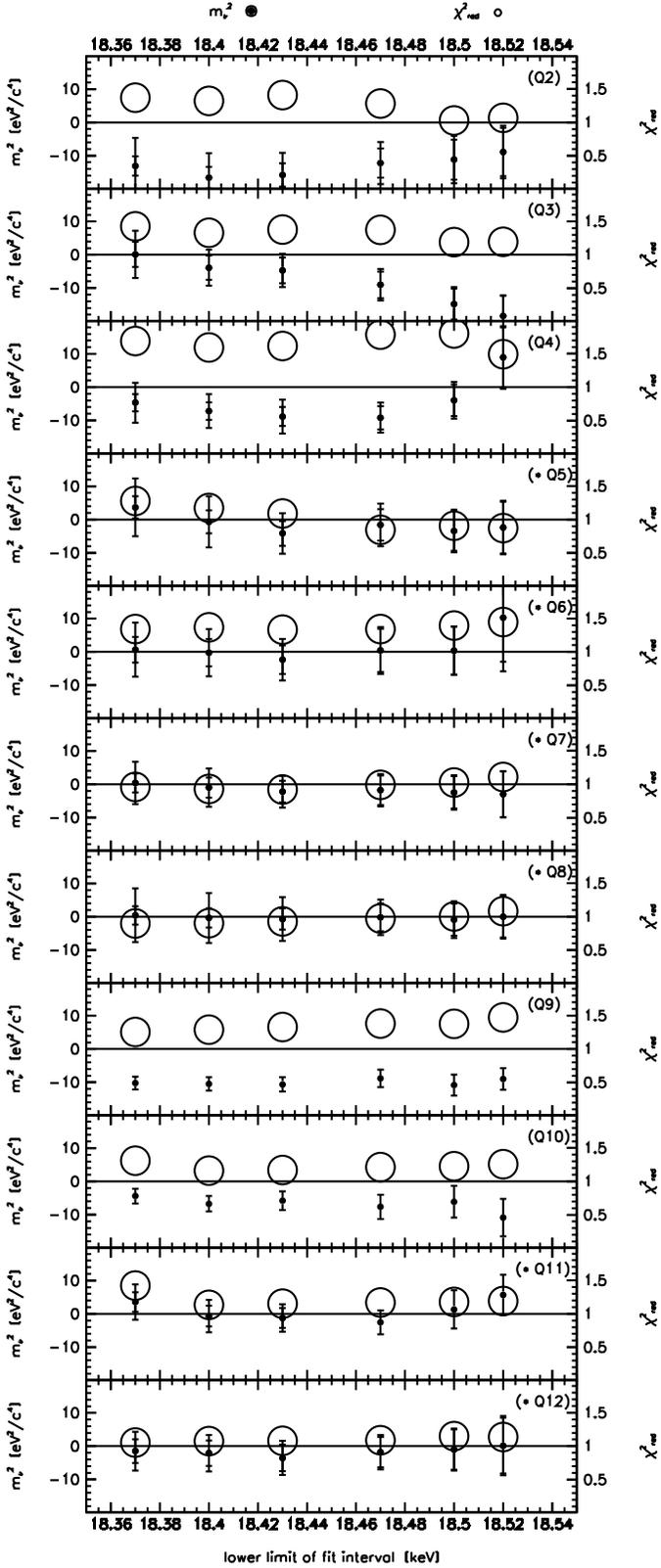, width=0.5\textwidth}
\caption{Fit results of ${m^2(\nu_e)}$ (data points, left scale) and reduced 
${\chi}^2$ (circles, right scale) for runs Q2 to Q12 in dependence of
the lower limit of the fit interval. The upper limit was always 18.6~keV. 
The inner error bars correspond to the statistical, the outer to the total 
uncertainty (except for Q9,Q10). The measurements used in the final analysis 
are marked by a star.} 
\label{figure13}
\end{figure}

\begin{figure}
\epsfig{file=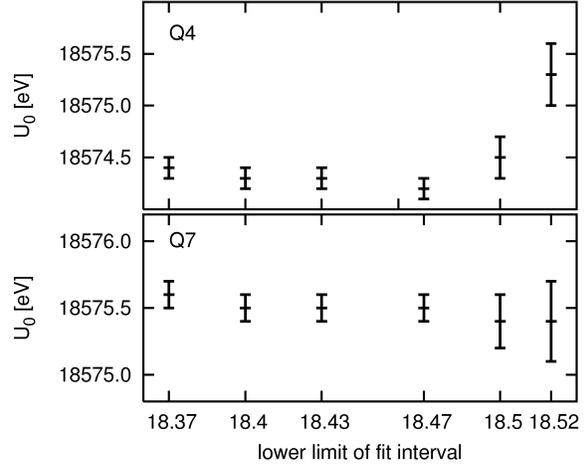, width=0.5\textwidth}
\hspace{-1cm}
\caption{Retarding voltages at which the fit finds endpoint values for runs 
Q4 and Q7 in dependence of the fit interval.}
\label{figure14}
\end{figure}

For all runs Q2 to Q12 fits were performed on data intervals of
different spectral extension. Fig.~\ref{figure13} shows the resulting 
$m^2(\nu_e)$ as function of the lower cut off of the accepted data interval. 
In some cases we observe small but still significant negative $m^2(\nu_e)$ 
values. For runs Q4 and Q7 the corresponding fit values for $E_0$ are 
shown separately in Fig.~\ref{figure14}. Their variation relative
to that of $m^2(\nu_e)$ reflects grosso modo the correlation (\ref{gln_corr}).

\begin{figure}{
\epsfig{file=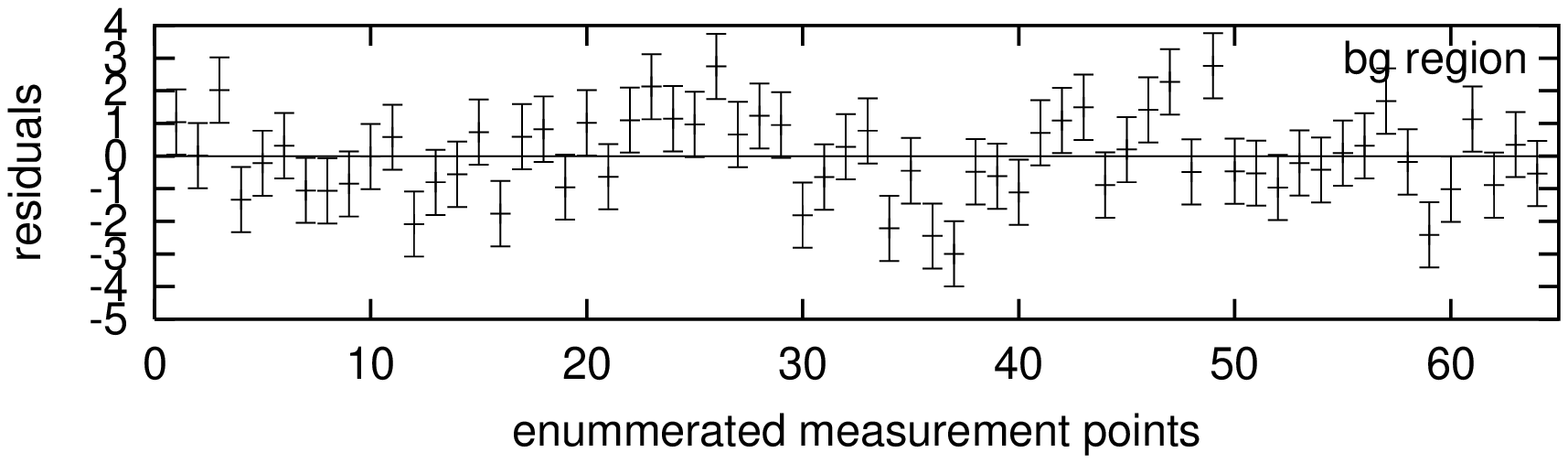, width=0.15\textwidth, angle=-90}
\epsfig{file=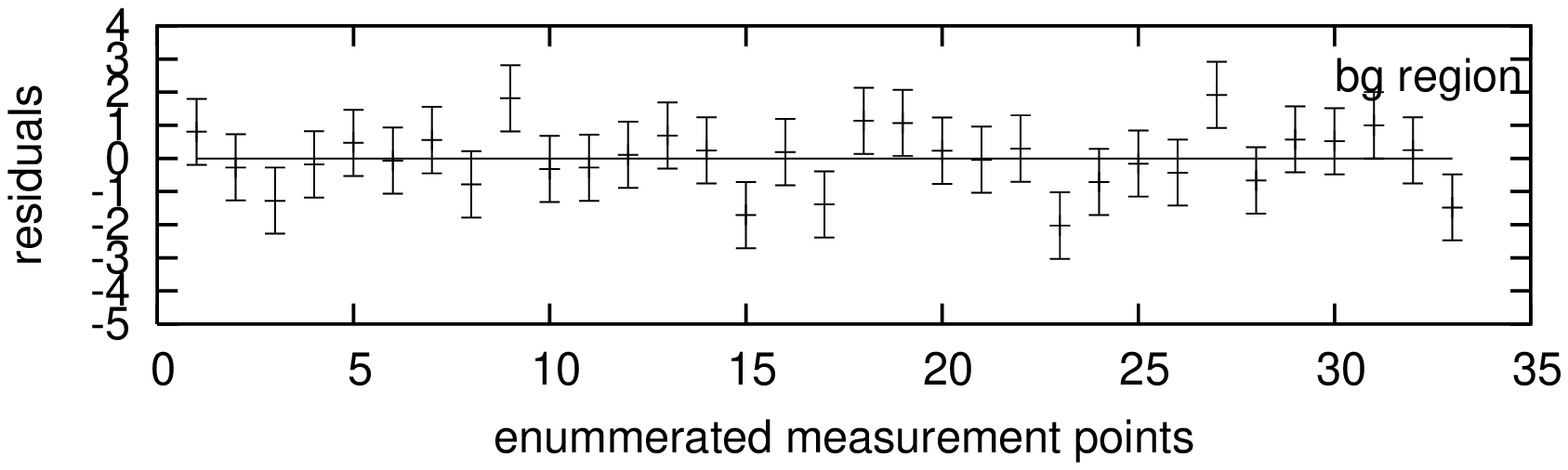, width=0.15\textwidth, angle=-90}  
}
\caption{Residua of fits for runs Q4 (upper part) and Q7 (lower part). 
The measurement points are enummerated to avoid misleading of the eyes due 
to unequal distances of the points. From left to right they correspond from 
200~eV below to 100~eV above the endpoint $E_0$.} 
\label{figure15}
\end{figure}

Fig.~\ref{figure13} shows in addition the reduced $\chi^2$ values for all 
fit intervals. In cases of negative or unstable $m^2(\nu_e)$ values they 
usually exceed 1, whereas they lie in the optimal range for data sets with 
straight $m^2(\nu_e)\approx 0$ fit results. This is seen more clearly in
Fig.~\ref{figure15} by the comparison of the residua of the straight
data set Q7 and the somewhat distorted one Q4.

\subsection{Run selection for the final evaluation}

The high statistics of individual runs in phase II reveals small
systematic differences between their results, which have been
discussed in the preceding section. Since they fall into two clear-cut
classes, clean ones and those with residual problems, we may select
only the former ones for the final evaluation in order to minimize
systematic deviations and uncertainties and to arrive at our optimum
overall result. This procedure might be considered questionable if the
selection were based on a negative $m^2(\nu_e)$ result alone; because
it might then introduce a bias. But we have clearly shown that
negative and fluctuating $m^2(\nu_e)$ values (the latter as function of
the data interval) occur at unstable running conditions, in particular
with respect to background.  Hence we can well justify our choice. In
this sense we retain runs Q5 to Q8, Q11 and Q12 for the final analysis. 
It should be noted, that only starting with run Q5 the very important 
method of applying high frequency pulses to one electrode at the detector 
side of the spectromter has been used in order to clean the spectrometer 
from possibly stored particles and to stabilize the background rate.

In detail Q2 had to be rejected because we had prepared an obviously
too thick source from which we learned about the unexpected charge up
effect. Here it spread over 6~eV, which is probably too much to be
corrected for safely. The other 4 rejected runs suffered from small
residual spectral anomalies of the data discussed above. Only in run
Q4 we observed an anomaly whose signature was compatible with a
Troitsk anomaly, i.e. a step rise in the integral spectrum. In this
case one might follow the Troitsk procedure of analysis. It consists
of fitting 2 additional parameters for position and size of the step
to the data. For distinct steps lying not too close to the endpoint
the additional parameters decorrelate sufficiently from the mass
parameter such that the fit yields reasonably stable $m^2(\nu_e)$ values close
to zero \cite{lobashev99}. This is the case for Q4 \cite{Weinheim99} (and
actually only for Q4). However, we have decided to retain from this ad
hoc procedure, since the step effect is neither stable nor properly
understood (see also section 6.2). This way we facilitate at least
the discussion of systematic uncertainties below.

\subsection{Joint analysis of selected runs} 
\begin{figure}
    \epsfig{file=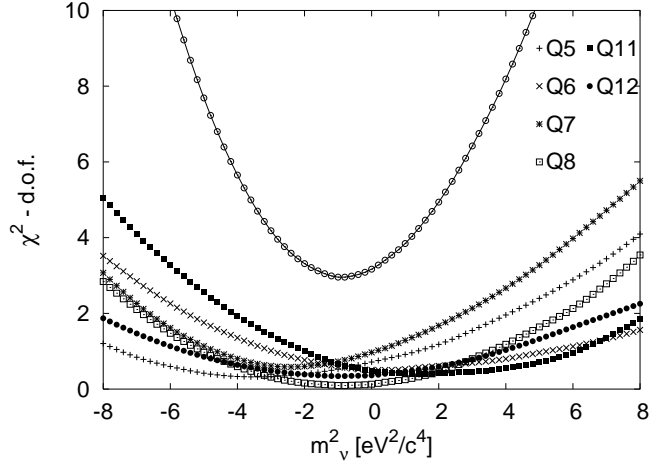, width=0.35\textwidth, angle=-90} 
\caption{Determination of $m^2(\nu_e)$ from the final data set. 
  Shown are the $\chi^2$ plots for the parameter $m^2(\nu_e)$ for the
  single data sets (different symbols) and the sum (open circles), which 
  corresponds to the fit of the total data set. The fit intervals are 
  restricted to a lower limit of 70~eV below $E_0$.}
\label{fig:chi}
\end{figure}

The data from the selected runs Q5 to Q8, Q11 and Q12 cannot be simply
summed up for a single fit, since they have been collected at somewhat
different conditions with respect to source strength, accepted solid
angle, choice of measuring points etc. We also have to face slight
changes of the fitted endpoint value beyond the statistical limit of
order 100 meV, since we cannot guarantee the stability of our HV
equipment to that level over years.  Actually $m^2(\nu_e)$ is the only 
parameter expected to approach one and the same value in any correct data 
set. Therefore, we have performed a joint fit of the full data set with 
respect to only this parameter by the following procedure. We have first 
fitted each of the selected runs separately with respect to amplitude,
background, and endpoint, and have calculated its $\chi^2$ as a function of
the common parameter $m^2(\nu_e)$. The six $\chi^2$ curves were then
added up to form a global $\chi^2$ curve (Fig.~\ref{fig:chi})
from which the final $m^2(\nu_e)$ fit result and its statistical error
are determined. This procedure is equivalent to a common fit of all
six data sets with 3$\cdot$6+1=19 free parameters; but it converges
much faster, since it makes proper use of the fact that each subset
depends only on three individual and one common parameter.

\subsection{Uncertainties of input parameters}
\label{uncer}
For most of the input parameters entering our final fit, we adapt the
same values and systematic uncertainties as chosen before in ref.s 
\cite{Weinheim99,Bonn}. For the prompt neighbour excitation in solid
T$_2$, however, we present in addition a new, critical treatment
below. Some uncertainties have been quoted in section \ref{sec:fitres} 
already. The others are discussed in the following.

\textbf{Final states of the daughter molecule:} We use the most recent
calculation by Saenz et al. \cite{Saenz} which have been calculated for 
gaseous T$_2$ with fully satisfactory precision. In solid T$_2$ the 
excitation energy of higher excited final states shifts up slightly
with respect to the ground state of (T$^3$He)$^+$. The effect has been
estimated by A.~Saenz \cite{saenz_private} and is considered here with a
correction of 0.8~eV for the second electronically excited state group and 
with 1.4~eV for the third one.
This correction is also considered fully as uncertainty.

\textbf{Energy loss in the T$_2$ film:} Spectrum and cross section of
energy loss have been discussed already in section 5.2. The relative
uncertainty of the latter is 5.4\% which is added in quadrature to
the one of the column density of individual runs according to Table 1.

Moreover, we consider a continuously growing coverage of the source by
0.3 monolayers of H$_2$ per day. This number has been obtained in two
independent ways \cite{BornL}: \\
(i) Ellipsometric determination of the
source thickness at the beginning and end of each run has revealed a
growth of the total thickness of 0.14 monolayers per day in the
average. Subtracting a T$_2$ loss of 0.17 monolayers/day as calculated
from the apparent lifetime of the source of 400 days, yields a growing
coverage by 0.31 monolayers/day. \\
(ii) Evaluation of data subsets from
fresh and older sources shows a significant trend towards negative
$m^2(\nu_e)$ values for the older ones. A coverage growth of 0.29
monolayers/day raises this dependence. The uncertainty of both results
is clearly larger than their difference. Therefore the correction by
0.3 monolayers/day is also considered fully as uncertainty.

\textbf{Neighbour excitation:} The prompt excitation of \\
neighbours next 
to a decaying T$_2$ molecule has been estimated by Kolos in sudden
approximation \cite{Kolos}. The effect is due to the local relaxation
of the lattice following the sudden appearance of an ion. A rigorous
calculation of final states of the surrounding electron cloud is still
missing. Therefore, it is difficult to assign a proper uncertainty to
Kolos' estimated excitation probability of $P_{\rm ne}=0.059$ with a mean
energy of 14.6~eV. The latter number applies to the excitation
spectrum of free hydrogen molecules. It seems reasonable to raise this
number by the same 1.5~eV by which the energy loss spectrum of
electrons is shifted upwards (compare section \ref{sec:fitres}). In the same
sense, the corresponding reduction of the total inelastic cross
section by 13\% \cite{Aseev} has been applied also to $P_{\rm ne}$ in our
former standard analysis. Another reduction of $P_{\rm ne}$ by 11\% has
been accounted for the observed porosity of our shock condensed films
(see section 3.1), yielding finally $P_{\rm ne}=0.046$. This number has
also been used in \cite{Weinheim99}, although it has been composed
from slightly different factors. Since the shifts in excitation energy
and probability, applied to Kolos' calculated values, are based but on
qualitative, plausible arguments, they had entered also fully the
systematic uncertainty.

\begin{figure}
  \hspace{-1.5cm}
    \epsfig{file=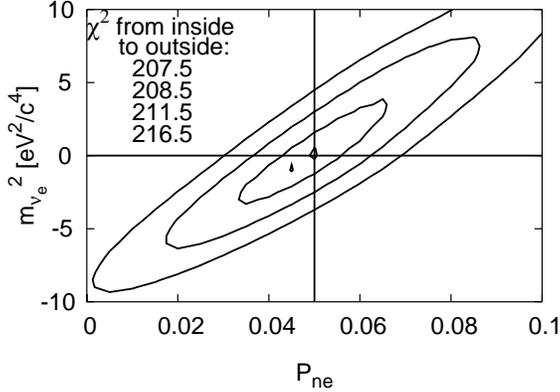, width=0.45\textwidth, angle=-90}
\caption {Shown are contour plots of ${\chi}^2(P_{\rm ne},m^2(\nu_e))$ at 
1${\rm \sigma}$, 2${\rm \sigma}$ and 3${\rm \sigma}$ around its minimum value 
at $(0.05,0)$.}
\label{fig:contour}
\end{figure}

The large, consistent data basis available now gives us a handle to
try a self-consistent determination of $P_{\rm ne}$ by treating it as an
additional free parameter in a joint analysis of a large interval
comprising data down to 170~eV below $E_0$ where the inelastic
components really matter. 
Fig.~\ref{fig:contour} shows the resulting
$\chi^2$ contour plot in the \\
$(P_{\rm ne},m^2(\nu_e))$ plane. Its minimum
lies at $P_{\rm ne}=0.050 \pm 0.016$ and $m^2(\nu_e)=(0\pm 3.3)$ eV$^2$/c$^4$,
a very satisfactory result, indeed. It confirms our former, estimated value 
of $P_{\rm ne}=0.046$ within errors and lifts the former tendency towards 
slightly negative $m^2(\nu_e)$ values for the entire data set (see below).

\begin{figure}
  \epsfig{file=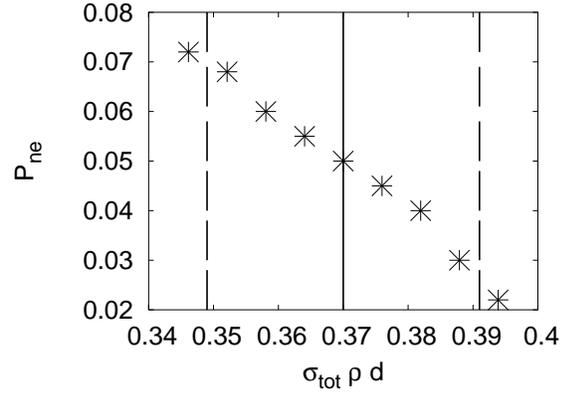, width=0.30\textwidth, angle=-90}
\caption{Correlation between neighbour excitation probability $P_{\rm ne}$ and 
$\sigma_{\rm tot} \rho d$. Each point corresponds 
to the centre of a contourplot like in Fig.~\ref{fig:contour}, calculated 
for particular values of $\sigma_{\rm tot} \rho d$ within its 1${\rm \sigma}$ 
uncertainty interval $(0.370\pm 0.023)$, indicated by the vertical lines.}
\label{fig:corr}  
\end{figure}

Still this fit value of $P_{\rm ne}$ is subject to a quite sizeable
systematic uncertainty of $\pm 0.022$. It stems from its strong
correlation to the energy loss in the film due to their similar effect on 
the spectrum. The uncertainty for energy losses are a combination of the 
uncertainty of the determination of the film thickness and the measurement 
of the total cross section $\sigma_{\rm tot}$. 
The averaged uncertainty in measuring the film thickness for all accepted 
runs is 3\% and the uncertainty of $\sigma_{\rm tot}$ is 5.4\%.
In order to determine the correlation of $P_{\rm ne}$ and $\sigma_{\rm tot} 
\rho d$~\footnote{$\sigma_{\rm tot} \rho d$ is a measure for the scattering 
probability in the tritium film.} we have calculated the ${\chi}^2$ contour 
plot of Fig.~\ref{fig:contour} for different values of 
$\sigma_{\rm tot} \rho d$ and located it's minimum in the 
$(P_{\rm ne}, m^2_{\nu_e})$- space each time. The 
corresponding $(P_{\rm ne},\sigma_{\rm tot} \rho d)$- pairs are plotted in 
Fig.~\ref{fig:corr}. We see an almost straight anticorrelation of the two 
energy loss contributions. The correlation transfers the uncertainty of 
$\sigma_{\rm tot} \rho d$ directly into one of $P_{\rm ne}$ as indicated by 
the bars in Fig.~\ref{fig:corr}. $m^2(\nu_e)$ and ${\chi}^2$ are rather 
insensitive to this exchange of $\sigma_{\rm tot} \rho d$ and $P_{\rm ne}$. 
In particular, we cannot fix $\sigma_{\rm tot} \rho d$ separately by the fit 
alone better than by external input. 

\textbf{Self-charging of T$_2$ film:} It has been found that the film
charges up within 30~min to a constant critical field strength of 62.6~MV/m 
\cite{auflad}. It results in a linearly
increasing shift of the starting potential of $\beta$ particles
throughout the film, reaching about 2.5~V at the outer surface. We
have assigned a conservative systematic uncertainty of $\pm 20\%$ to
that slope.

\textbf{Backscattering and detector efficiency:} Both effects are small
and can be accounted for by the linear correction factors 
(\ref{alphaback},\ref{fbackcorr}) and (\ref{fdcorr},\ref{alphad}) given 
above. They may just be contracted into a single correction factor
\begin{equation}
f_{\rm back corr}\cdot f_{\rm d corr}\approx 1+(\alpha_{\rm
  back}+\alpha_{\rm d})(E+eU).
\label{singcorrec}
\end{equation}
In (\ref{alphad}) we have already assigned to $\alpha_{\rm d}$ a conservative
uncertainty of $\pm 2\cdot 10^{-5}/$eV, which is large enough to cover
a residual uncertainty of $\alpha_{\rm back}$ as well.

\begin{figure}
   \epsfig{file=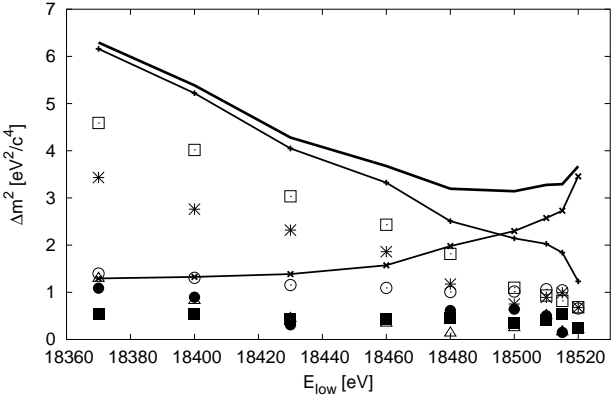, width=0.30\textwidth, angle=-90} 
\caption{Individual and quadratically summed up uncertainties of $m^2(\nu_e)$ 
for the joint data set, calculated for different lengths of the data interval. 
The open squares and stars give the contribution of energy losses in the 
tritium film, the open circles the neighbour excitation, the filled squares 
the self-charging effect, the filled circles the final states and the 
open triangles the detector efficiency. The line with simple crosses shows 
the sum of systematic uncertainties in dependence from the fit interval and 
the line with stars gives the corresponding statistical uncertainty 
(growing). The upper most line gives the quadratically summed up values and 
has its minimum for 18500~eV.}
\label{fig:syssum}
\end{figure}

\subsection{Systematic uncertainty of 
${\boldsymbol m}^{\boldsymbol 2}({\boldsymbol \nu}_{\boldsymbol e})$}
In the standard analysis the systematic uncertainty of $m^2(\nu_e)$ is
calculated from those of the external input parameters as follows:
Each input parameter is shifted from its best value by its uncertainty
and the fit to a particular data set is repeated.  The resulting shift
of the $\chi^2$ minimum with respect to $m^2(\nu_e)$ is then taken as
the corresponding systematic uncertainty of $m^2(\nu_e)$.  

The systematic uncertainty of the joint data set is evaluated in the standard 
analysis by the same procedure. Fig.~\ref{fig:syssum} shows for the joint 
data set the individual contributions as well as the quadratically summed up
uncertainties calculated for different lengths of the data interval.
Obviously the statistical uncertainty decreases with the length of the
interval whereas the systematic increases. The total, combined
uncertainty attains a flat minimum of ${\rm \Delta} m^2_{\rm tot} = 
3.04$~eV$^2$/c$^4$ for a lower cut off at $E_{\rm low}=18500$~eV which is 
regarded the optimum interval, therefore.

Summarizing, the standard analysis with the external input parameter of 
neighbour excitation probability $P_{\rm ne}=(0.046\pm 0.013)$
yields for the optimal data interval the result
\begin{eqnarray}
m^2(\nu_e)_{\rm standard} = (-1.2\pm 2.2_{\rm stat}\pm 
2.1_{\rm syst})~\textrm{eV}^2/\textrm{c}^4; \nonumber \\
\chi^2/{\rm d.o.f.}=208/195.
\label{mstand}
\end{eqnarray}
Here the uncertainties of the externally chosen, independent
parameters $P_{\rm ne}$ and $\sigma_{\rm tot} \rho d$ contribute to that of 
$m^2(\nu_e)$ by an amount of
\begin{equation}
\delta m^2(\nu_e)_{\rm syst, \rm ext}(P_{\rm ne},\sigma_{\rm tot} \rho d)=
1.59~\textrm{eV}^2/\textrm{c}^4. 
\label{deltasys}
\end{equation}
Using the self consistently fitted $P_{\rm ne}$ value instead, we
have to take into account that its systematic uncertainty of $\pm 0.022$ is
anticorrelated to the corresponding uncertainty of $\mp 0.023$ of 
$\sigma_{\rm tot} \rho d$ according to Fig. \ref{fig:corr}. Their combined 
action on $m^2(\nu_e)$ has to be calculated, therefore. Moreover, we have to 
consider the statistical uncertainty $\delta P_{\rm ne,\rm stat}=\pm 0.016$ 
which results from the fit in Fig.~\ref{fig:contour}. Added in quadrature to 
the systematic contribution we obtain from the self consistent analysis for 
the optimum interval $E>18500$~eV a combined systematic uncertainty of
\begin{equation}
\delta m^2(\nu_e)_{\rm syst, \rm selfcons.}(P_{\rm ne}, 
\sigma_{\rm tot} \rho d)= 1.58~ \textrm{eV}^2/\textrm{c}^4. 
\label{deltasys2}
\end{equation}
The marginal reduction compared to (\ref{deltasys}) would not really be worth 
the effort. Rather we emphasize that it determines for the first time
the probability of neighbour excitation from the data themselves  and confirms,
moreover, the qualitative estimation of correction factors applied
earlier to Kolos' original calculation of $P_{\rm ne}$ \cite{Kolos}.

In addition the slight shift of the central value of $P_{\rm ne}$ from 0.046 
in the standard analysis to 0.05 in the self-consistent one causes a
corresponding shift of the $m^2(\nu_e)$ fit value of the final
result still closer to zero

\begin{eqnarray}
m^2(\nu_e)=(-0.6\pm 2.2_{\rm stat}\pm 2.1_{\rm syst})~\textrm{eV}^2/\textrm{c}^4\nonumber \\
\chi^2/{\rm d.o.f.}=208/194
\label{gln_afmerg} 
\end{eqnarray}

\section{Discussion of results}
\subsection{Experimental $\boldsymbol \beta$ spectrum}

\begin{figure}
 \epsfig{file=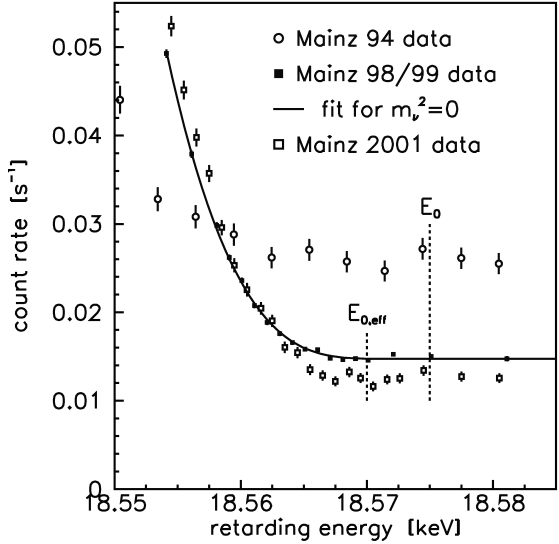, width=0.45\textwidth}
\caption{Averaged count rate of the 98/99 data (filled squares) with fit for 
$m^2(\nu_e)=0$ (line) and the 2001 data (open squares) in comparison with 
previous Mainz data from phase~I (open circles) plotted as function of the 
retarding potential near the endpoint $E_0$.}
\label{fig:spectren}
\end{figure}

We start the discussion by taking a look at the measured spectra in
the vicinity of the endpoint (Fig. \ref{fig:spectren}).  The bulk of
phase II data has been obtained in runs Q5 to Q8 under almost
identical conditions and may hence be composed here to a single spectrum 
(full squares). The open squares represent
runs Q11 and Q12. The rate is slightly higher and the background still lower 
than in the Q5 to Q8 runs. For comparison we also show the last
spectrum from phase I taken in 1994 (open circles) \cite{Backe96,Barth97}. 
The full curve is a fit to the Q5 to Q8 data with $m^2(\nu_e)$ fixed to zero, 
looking perfect. Already 5~eV below the effective endpoint the spectrum rises
distinctly from the background, excluding prima vista any larger
neutrino mass. (The shift of the effective endpoint from the true one
is obtained from an average over the ro- vibrational excitation of the
daughter molecule, over the transmission function, and over the source
charging). Moreover, these summed up data exclude safely any steady
spectral anomaly close to the endpoint on the level of 1~mHz; this
corresponds to about 10$^{-12}$ of the total decay rate of the source.

\subsection{Troitsk anomaly}
\begin{figure}
    \epsfig{file=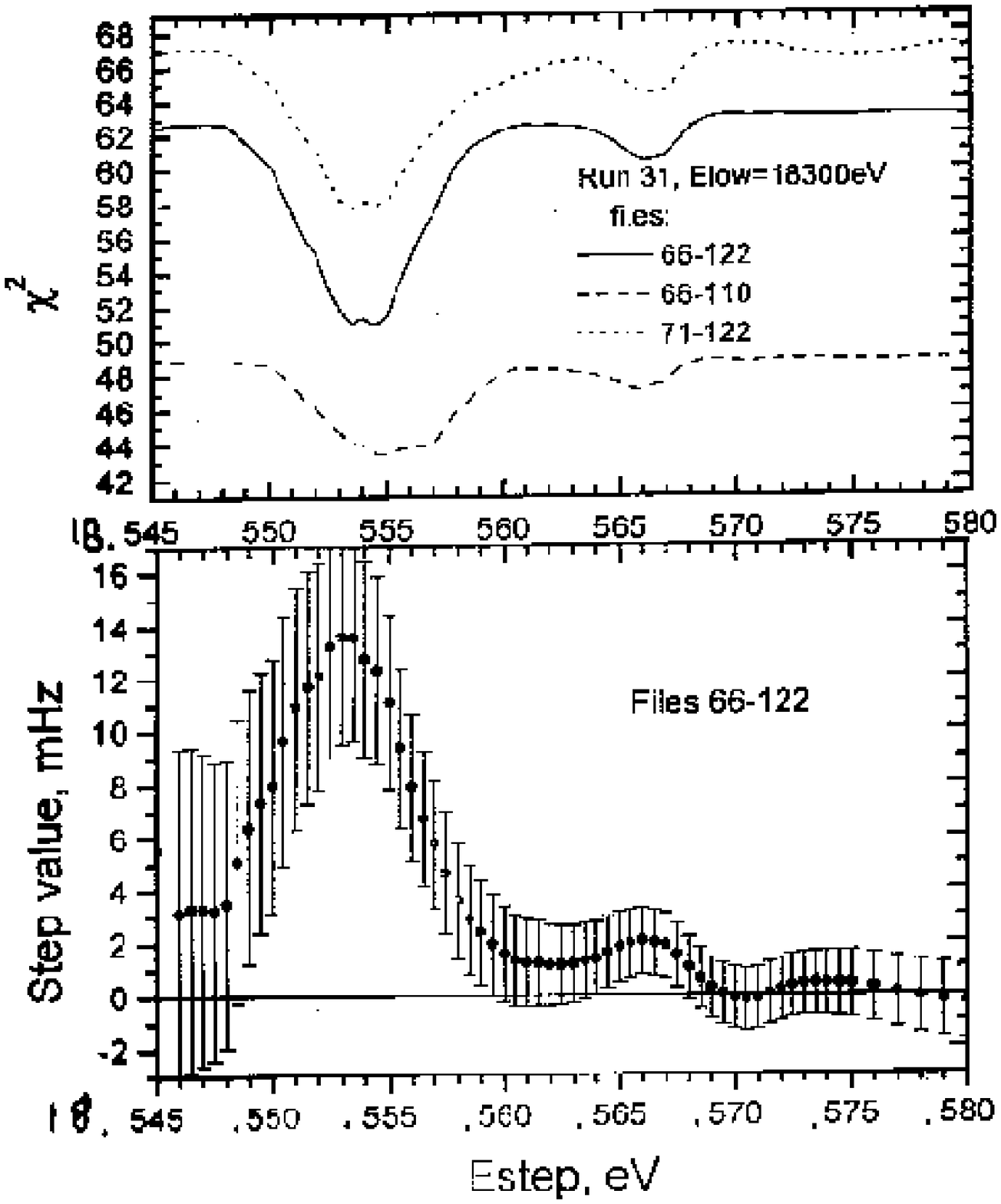, width=0.5\textwidth}
    \epsfig{file=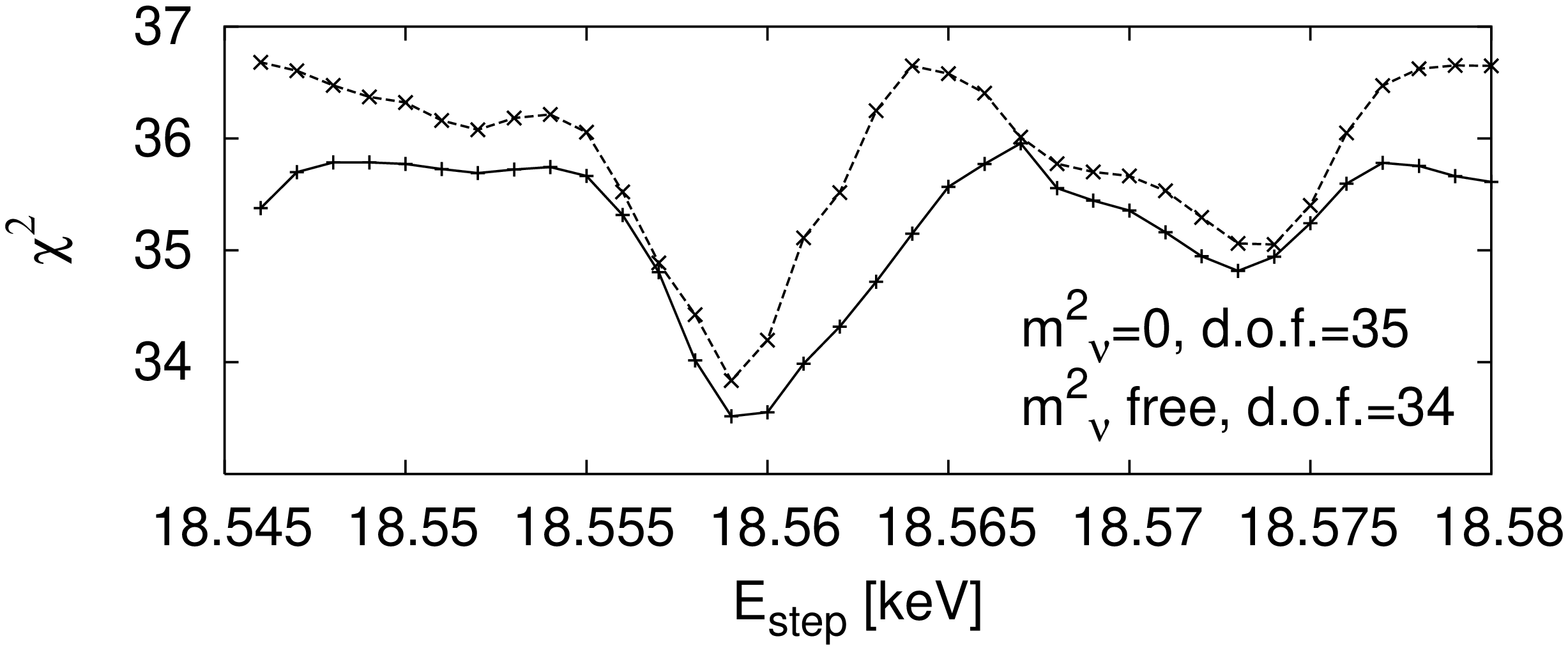, width=0.20\textwidth, angle=-90} 
\caption{Search for a step like anomaly in parallel measurements in december 
2000. The upper and the middle graph show the analysis of the Troitsk data 
\cite{lobashev01} by adding position and amplitude of the step as free 
parameters. 
The upper graph shows the resulting drop of $\chi^2$, the lower one the 
fitted step amplitude, both as function of the step position. A very 
significant signal appears around 18553~eV. In contrast the corresponding 
$\chi^2$ plot for the Mainz data (lower graph) is shown, fluctuations are 
insignificantly by only 2 units.} 
\label{fig:coinci}
\end{figure}

The absence of any anomaly in the summed up spectrum does not exclude
necessarily a step like Troitsk anomaly which would fluctuate in position and
amplitude. It might be washed out in the summed up data.  Indeed it
has been observed to be a fluctuating effect, for some time hinting to
a half year period even \cite{lobashev99}. In December 2000, however, it
appeared as sudden outburst \cite{lobashev01}. In this period Q10
was running at Mainz in parallel. Fig. \ref{fig:coinci}
shows the analysis of both runs with respect to the appearance of a
step in the integral spectrum, i.\,e.\, a line in the original
spectrum. To that end one fits the spectrum to the data under the
assumption of an additional sharp line of free amplitude at a
particular position. The upper plot shows the
course of $\chi^2$ as function of the line position for the Troitsk
data.  A very significant minimum is observed at 18553~eV indicating a
line, (or step, respectively) with an equally significant amplitude of
13~mHz (middle plot). The corresponding
$\chi^2$ plot for the parallel run at Mainz with similar sensitivity 
shows but fluctuations of statistical size (lower plot). Hence
speculations that the Troitsk anomaly might be due to a fluctuating
presence of dense neutrino clouds \cite{lobashev99} are disproved.  
Rather it has to be attributed to instrumental effects, as pointed out 
already in section 4.2 (see also \cite{CKraus} and \cite{Kraus}).

\subsection{$\boldsymbol m^{\boldsymbol 2}({\boldsymbol \nu}_{\boldsymbol e})$ 
 result and upper limit of 
$\boldsymbol m({\boldsymbol \nu}_{\boldsymbol e})$}
\begin{figure}
\epsfig{file=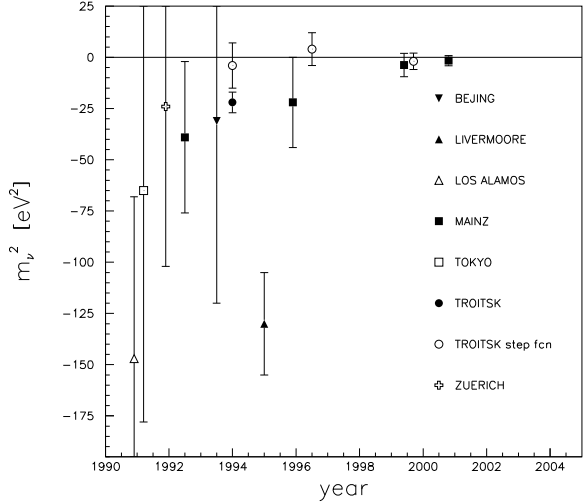, width=0.5\textwidth}
\caption{Published results of squared neutrino mass values $m^2(\nu_e)$ 
obtained from tritium decay since 1990. The already finished experiments at 
Los Alamos, Z\"urich, Tokyo, Beijing and Livermore  
\cite{losala,zurich,tokyo,beijing,liver} 
used magnetic spectrometers, the experiments at Troitsk and Mainz are using 
MAC-E-Filters (as described before).}
\label{fig:results}
\end{figure}

From the two alternative choices of the
neighbour excitation probability we settle on the self-consis\-tently
determined one for reasons given in sections 5.6 and 5.7. Hence 
(\ref{gln_afmerg}) is our final experimental result
on the observable $m^2(\nu_e)$. As compared to our ealier communicated
result \cite{Bonn} it has improved in three respects: 
\begin{enumerate}
\item The statistical uncertainty has been diminished further by 
0.3~eV$^2$/c$^4$. 
\item The systematic uncertainty has been better founded with respect
  to the questionable neighbour excitation probability (but remained
  unchanged in size). 
\item  The central value has moved further up from -1.6 eV$^2$/c$^4$ to
  -0.6~eV$^2$/c$^4$ and has lost by now any touch of being unphysical in
  view of the error bars. 
\end{enumerate}
The progress in the observable $m^2(\nu_e)$ of this final Mainz result
as compared to the most sensitive earlier experiments using momentum analysing 
spectrometers approaches 2 orders of magnitude (Fig. \ref{fig:results}). 
The Troitsk group communicated similar numbers \cite{lobashev03}  
($m^2(\nu_e)=(-2.3\pm 2.5_{\rm stat}\pm 2.0_{\rm syst})~
\textrm{eV}^2/\textrm{c}^4$), but there is an important difference. The 
Troitsk group needs to correct for the observed anomaly by adding 
phenomenologically a sharp line with free position and size to the $\beta$ 
spectrum without including a systematic correction for this approach.
Without this correction the fit would charge this effect on $m^2(\nu_e)$ and 
drive it negative as discussed above already.
Since phenomenology and origin of the anomaly are barely known, this 
procedure is not obvious and it is difficult to assign a proper systematic 
uncertainty to this correction. Up to now the Troitsk group has not 
considered in its result any systematic uncertainty of this correction.

If we move the central value
of $m^2(\nu_e)$ to zero and calculate from there the upper mass limit
at 95\%~C.L., then we obtain the so-called
sensitivity limit. It lies for both evaluations (\ref{mstand}) and
(\ref{gln_afmerg}) at

\begin{equation}
m_{\rm sens.~lim.}(\nu_e) = 2.4~\textrm{eV}/\textrm{c}^2 \quad 
(95\%~{\rm C.L.}).
\label{hoch}
\end{equation}

Since the actual $m^2(\nu_e)$ values are slightly negative we derive from them an upper limit by help of the unified approach \cite{unified},
recommended by the particle data group. This yields in case of the
result (\ref{mstand}) from the standard analysis

\begin{equation}
m(\nu_e)<2.2~\textrm{eV}/\textrm{c}^2 \quad (95\%~{\rm C.L.})
\label{niedrig}
\end{equation}

which agrees with the latest communicated value \cite{Bonn}. Our
preferred result (\ref{gln_afmerg}), however, calculated with a
self-con\-sistent $P_{\rm ne}$ value, yields a slightly higher limit

\begin{equation}
m(\nu_e)<2.3~\textrm{eV}/\textrm{c}^2 \quad (95\%~ \rm C.L.)
\label{gln_mainzlimit}
\end{equation} 
since the respective $m^2(\nu_e)$ value lies still closer to zero. The 
increasing reduction of the upper limit below the sensitivity limit as 
function of an increasing negative $m^2(\nu_e)$ value, is but a dubious gift 
of the unified approach, accompanied by an increase of the probability that 
the result suffers from unidentified systematic errors. We quote
(\ref{gln_mainzlimit}) instead of (\ref{hoch}) or (\ref{niedrig}) as our 
final upper mass limit, because it stems
from a more consistent analysis of the data on one hand and conforms to
the recommended unified approach on the other. Anyway, their difference is 
marginal.

\section{Conclusion and outlook}
Phase II of the Mainz neutrino mass experiment started in 1995 with
substantial improvements regarding the frozen T$_2$ source as well as
background rejection in the $\beta$ transport channel and in the
electrostatic spectrometer. This enabled running at a 10 times better
signal to background ratio from 1997 on. Thereafter,
a number of side experiments yielded: 
\begin{enumerate}
\item A detailed study and suppression of the unexpected and disturbing
dewetting of the T$_2$ film from the substrate \cite{Fleisch2,Fleisch3}. 
\item The discovery, quantification and theoretical explanation of source 
charging \cite{auflad}. 
\item A determination of the energy loss spectrum of $\beta$ particles
in solid T$_2$ \cite{Aseev}. 
\item Phenomenological studies and suppression of background mechanisms
in MAC-E Filters. 
\end{enumerate}
They formed the basis for a satisfactory control and reduction of systematic
uncertainties in parallel to the statistical one. 

Data taking on the search for the neutrino mass covered the years 1998 to 
2001 and yielded the so far narrowest limit on the observable $m^2(\nu_e)$ 
of $(-0.6 \pm 3.0)~\textrm{eV}^2/\textrm{c}^4$ from which an upper limit
$m(\nu_e)<2.3~\textrm{eV}/\textrm{c}^2$ (90\% C.L.) is derived.

The discovery of neutrino oscillations at squared mass differences of
${\rm \Delta} m^2_{ij}\leq 0.05 {\rm eV}^2/{\rm c}^4$
\cite{SuperK,Davis,SK,original,Gallex,Sage,SNO,SNO2,Kamland,Garcia} allows
furthermore to apply the same upper limit to all three neutrino
flavours as reference value in particle and astrophysics. Clearly
there is a burning interest to improve this limit further in order to
check cosmological models more sensitively by laboratory results on
one hand and to decide the alternative between degenerate and
hierarchical neutrino masses on the other. In this respect a recent
paper on the final data from the Heidelberg Moscow experiment is
giving a first indication; it reports on a 4${\rm \sigma}$ signal of
neutrinoless double $\beta$ decay in $^{76}\textrm{Ge}$
\cite{Klapdor04}. Explained by an exchange of a massive Majorana neutrino, 
this signal would give a mass in approximate limits 
$0.1\leq m_{ee}~\textrm{c}^2/\textrm{eV} \leq 0.9$ (95\%~C.L.).

Obviously our present experiment has exhausted its potential by now,
almost 20 years after it has been first conceived. First plans to build 
either a rigorously enlarged MAC-E-Filter \cite{Barthetal98}, or a bent 
variant offering in addition a highly resolving differential energy analysis 
\cite{Lobashev98} were proposed at an Erice meeting in 1997. A following 
paper treated in some detail the potential of a MAC-E-Filter of 5~m diameter 
and reported moreover on an implementation of a time of flight mode which
transforms a MAC-E-Filter from a high pass to a narrow band filter
with equally sharp slopes \cite{bonn99}. In 2000 the KATRIN collaboration
\cite{Katrin} formed proposing to build a large MAC-E-filter in
combination with a gaseous T$_2$ source at the site of the Forschungszentrum
Karlsruhe. It combines the expertise from the foregoing experiments
at Los Alamos \cite{Wilkerson}, Mainz and Troitsk with the strength of a 
national laboratory including expertise in handling large amounts of tritium. 
The present design aims at reaching within 3 years of measurement a precision 
of ${\rm \Delta} m^2(\nu_e)\approx 0.02 \textrm{eV}^2/\textrm{c}^4$
corresponding to a sensitivity limit of 0.2~eV/c$^2$ for the mass
itself. The experiment should be ready to go in 2008.

\section*{Acknowledgement}
The authors are grateful to the Deutsche Forschungsgemeinschaft, which
supported this experiment under contract no. Ot33 from its beginning
until 2000 and to the Bundesministerium f{\"u}r Bildung und
Wissenschaft which took over thereafter under contract no. 06Mz866I/5.
Our thanks also go to our frequent guests from Troitsk, V. Lobashev 
(supported by an Alexander von Humboldt award), as well as O. Kazachenko 
and N. Titov (supported by a visitor programme of the Deutsche 
Forschungsgemeinschaft) for their valuable contributions.

{}


\begin{thebibliography}{}
  
\bibitem{SuperK} S.~Fukuda and the Superkamiokande Collaboration, 
                 Phys.~ Rev.~Lett. {\bf 81}, 1562 (1998)

\bibitem{Davis} R.~Davis, Nucl.~Phys.~B (Proc. Suppl.) {\bf 48}, 284 (1996) 

\bibitem{SK} M.B.~Smy and the Superkamiokande Collaboration, Nucl. Phys. B 
             (Proc. Suppl.) {\bf 118}, 25 (2003) 

\bibitem{original} B.T.~Cleveland et al., Astrophys. J. {\bf 496}, 505, (1998) 

\bibitem{Gallex} T.A.~Kirsten and the GNO Collaboration, Nucl. Phys. B 
                 (Proc. Suppl.) {\bf 118}, 33 (2003) 

\bibitem{Sage} V.N.~Garvin and the SAGE Collaboration, Nucl. Phys. B 
               (Proc. Suppl.) {\bf 118}, 39 (2003)

\bibitem{SNO} A.L.~Hallin and the SNO Collaboration, Nucl. Phys. B 
              (Proc. Suppl.) {\bf 118}, 3 (2003) 

\bibitem{SNO2} Q.R.~Ahmad et al., Phys. Rev. Lett. {\bf 89}, 011301, (2002) 

\bibitem{Kamland} K.~Eguchi and the KamLAND Collaboration, Phys. Rev. Lett.  
                  {\bf 90}, 021802 (2003) 

\bibitem{Garcia} M.C.~Gonzalez-Garcia, Y.~Nir, Rev. Mod. Phys. {\bf 75}, 
                 345 (2003) 

\bibitem{Klapdor01} H.V.~Klapdor-Kleingrothaus, A.~Dietz, H.L.~Harney, 
                    I.V.~Krivosheina, Mod. Phys. Lett. {\bf A16}, 2409 (2001) 

\bibitem{Klapdor04} H.V.~Klapdor-Kleingrothaus, I.V.~Krivosheina, A.~Dietz, 
                    O.~Chkvorets, Phys. Lett. {\bf B586}, 198 (2004) 

\bibitem{Spergel} D.N.~Spergel et al., arXiv:astro-ph/0302209 (2003), 
                  Astrophys. J. (Suppl.) {\bf 148}, 175 (2003) 

\bibitem{Hannestad} S.~Hannestad, JCAP {\bf 0305}, 004 (2003), arXiv: 
                    astro-ph/ 0303076 (2003)

\bibitem{Barger} V.~Barger, D.~Marfatia, A.~Regre, arXiv: hep-ph/0312065 v2 
                 (2003) 

\bibitem{Allen} S.W.~Allen, A.C.~Fabian, S.L.~Fridle, arXiv: astro-ph/0306386 
                (2003) 

\bibitem{Robertson} R.G.H.~Robertson, D.A.~Knapp, Ann. Rev. Nucl. Sci. 
                    {\bf 38}, 185 (1988)

\bibitem{Holzschuh} E.~Holzschuh, Rep. Proc. Phys. {\bf 55}, 1035 (1992)

\bibitem{WilkersonRobertson} J.F.~Wilkerson and R.G.H.~Robertson in 
                             ``Current Aspects Of Neutrino Physics'', 
         edited by~D.O.~Caldwell (Springer, Berlin, Heidelberg, 2001), p. 39

\bibitem{Weinheim03} Ch.~Weinheimer in ``Neutrino Mass'', 
         edited by~G.~Altarelli and K.~Winter, Springer Tracts in Modern 
         Physics {\bf 190}, (Springer-Verlag, Berlin, Heidelberg, Germany, 
         2003), p. 25

\bibitem{Picard} A.~Picard, H.~Backe, H.~Barth, J.~Bonn, B.~Degen, Th.~Edling, 
         R.~Haid, A.~Hermanni, P.~Leiderer, Th.~Loeken, A.~Molz, R.B.~Moore, 
         A.~Osipowicz, E.W.~Otten, M.~Przyrembel, M.~Steininger, 
         Ch.~Weinheimer, Nucl. Instr. Meth. B {\bf 63}, 345 (1992) 

\bibitem{lobashev85} V.M.~Lobashev, P.E.~Spivac, Phys. Lett. B {\bf 460}, 
                     3305 (1985) 

\bibitem{Belesev} A.I. Belesev et al., Phys. Lett. B {\bf 350}, 263 (1995) 

\bibitem{Weinprom} Christian Weinheimer, Doct. Thesis, Mainz, Univ. (1993)

\bibitem{Weinheim93} Ch.~Weinheimer, M.~Przyrembel, H.~Backe, H.~Barth, 
         J.~Bonn, B.~ Degen, Th.~Edling, H.~Fischer, L.~Fleischmann, 
         J.U. Groo{\ss}, R.~Haid, A.~Hermanni, G.~Kube, P.~Leiderer, 
         Th.~Loeken, A.~Molz, R.B.~Moore, A.~Osipowicz, E.W.~Otten, 
         A.~Picard, M.~Schrader, M.~Steininger, Phys. Lett. B {\bf 300}, 
         210 (1993)

\bibitem{Backe96} H.~Backe, H.~Barth, A.~Beile, J.~Bonn, B.~Degen, 
         L.~Fleischmann, M.~Gundlach, E.W.~Otten, M.~Przyrembel, 
         Ch.~Weinheimer in Proc. 17. Int. Conf. on Neutrino Physics and 
         Astrophysics, Helsinki, Finnland, June 1996, World Scientific, 
         Singapore, p. 259 

\bibitem{Weinheim99} Ch.~Weinheimer, B.~Degen, A.~Bleile, J.~Bonn, 
         L.~Bornschein, O.~Kazachenko, A.~Kovalik, E.W.~Otten, Phys. Lett. B 
         {\bf 460}, 219 (1999) 

\bibitem{lobashev99} V.M.~Lobashev et al., Phys. Lett. {\bf B460}, 227 (1999)

\bibitem{Bonn} J.~Bonn, B.~Bornschein, L.~Bornschein, L.~Fickinger, B.~Flatt, 
               O.~Kazachenko, A.~Kovalik, Ch.~Kraus, E.W.~Otten, J.P.~Schall, 
               H.~Ulrich, Ch.~Weinheimer, Nucl. Phys. B (Proc. Suppl.) 
               {\bf 91}, 273, (2001)

\bibitem{endpoint} R.S.~Van Dyck, Jr., D.L.~Farnham, P.B.~Schwinberg, 
                   Phys. Rev. Lett {\bf 70}, 2888 (1993)  

\bibitem{Simpson} J.J.~Simpson, Phys. Rev. D {\bf 23}, 64 (1981) 

\bibitem{Repco} W.W~Repco, C.E.~Wu, Phys. Rev. C {\bf 28}, 2433 (1983) 

\bibitem{gardner} S.~Gardner, V.~Bernard, U.-G.~Mei$\ss$ner, Phys. Lett. B 
                  {\bf 598}, 188 (2004)

\bibitem{ste98} G.J.~Stephenson, T.~Goldman, Phys. Lett. B {\bf 440}, 89 (1998)

\bibitem{PART} Particle Data Group, Phys. Rev. D {\bf 66}, 321 (2002)

\bibitem{CKraus} Christine Kraus, Diploma Thesis, Mainz Univ. (2000)

\bibitem{Fackler} O.~Fackler, B.~Jeziorski, W.~Kolos, H.J.~Monkhorst, 
                  K.~Szalewicz, Phys. Rev. Lett. {\bf 55}, 1388 (1985) 

\bibitem{Saenz} A.~Saenz, S.~Jonsell, P.~Froehlich, Phys. Rev. Lett.  
                {\bf 84}, 242 (2000)

\bibitem{Kolos} W.~Kolos et al., Phys. Rev. A {\bf 37}, 2297 (1988)

\bibitem{Otten94} E.W.~Otten, Prog. Part. Nucl. Phys. {\bf 32}, 153 (1994)

\bibitem{Picard92} A.~Picard, H.~Backe, J.~Bonn, B.~Degen, R.~Haid, 
                   A.~Hermanni, P.~Leiderer, A.~Osipowicz, E.W.~Otten, 
                   M.~Przyrembel, M.~Schrader, M.~Steininger, Ch.~Weinheimer, 
                   Z.Phys. A {\bf 342}, 71 (1992) 

\bibitem{BornB} Beate Bornschein, Doct. Thesis, Mainz, Univ. (2000)

\bibitem{BornL} Lutz Bornschein, Doct. Thesis, Mainz, Univ. (2002) 

\bibitem{Kraus} Christine Kraus, Doct. Thesis, Mainz, Univ. (2004) 

\bibitem{Fleisch1} Lars Fleischmann, Doct. Thesis, Mainz, Univ. (1998) 

\bibitem{Fleisch2} L.~Fleischmann, J.~Bonn, B.~Degen, M.~Przyrembel, 
                   E.W.~Otten, Ch.~Weinheimer, P.~Leiderer, 
                   J. Low Temp. Phys. {\bf 119}, 615 (2000) 

\bibitem{Fleisch3} L.~Fleischmann, J.~Bonn, B.~Bornschein, P.~Leiderer, 
                   M.~Przyrembel, Ch.~Weinheimer, A.~Saenz, 
                   Euro. Phys. J. B {\bf 16}, 521 (2000)

\bibitem{Aseev} V.N.~Aseev, A.I.~Belesev, A.I.~Berlev, E.V.~Geraskin, 
                O.V.~Kazachenko, Yu.E.~Kuznetsov, V.M.~Lobashev, 
                R.P.~Ostroumov, N.A.~Titov, S.V.~Zadorozhny, Yu.I.~Zakharov, 
                J.~Bonn, B.~Bornschein, L.~Bornschein, E.W.~Otten, 
                M.~Przyrembel, Ch.~Weinheimer, A.~Saenz , Euro. Phys. J. D 
                {\bf 10}, 39 (2000) 

\bibitem{Leiderer} R.~Conradt, U.~Albrecht. S.~Herminghaus, P.~Leiderer, 
                   Physica B {\bf 194-196}, 679 (1994)

\bibitem{Kettig} Oliver Kettig, Dipl. Thesis, Mainz, Univ. (1994)

\bibitem{Azzam} R.M.A.~Azzam, N.M.~Bashara, Ellipsometry and polarized 
         light, North Holland Publ. Comp, Amsterdam, Netherlands (1988)

\bibitem{Kolos67} W.~Kolos, L.~Wolniewicz, J. Chem. Phys. {\bf 46}, 1426 
                  (1967) 

\bibitem{Silvera} I.~Silvera, Rev. Mod. Phys. {\bf 52}, 393 (1980) 

\bibitem{Fischer} Hermann Fischer, Dipl. Thesis, Mainz, Univ. (1980) 

\bibitem{Barth} Herbert Barth, Dipl. Thesis, Mainz, Univ. (1991)

\bibitem{Goldmann} Daphne Goldmann, Dipl. Thesis, Mainz, Univ. (1995) 

\bibitem{Katrin} A.~Osipowicz et al., arXiv:hep-ex/0109033

\bibitem{lobashev} V.M.~Lobashev, Troitsk, priv. commun.

\bibitem{Barth97} Herbert Barth, Doct. Thesis, Mainz, Univ. (1997)

\bibitem{Wein92} Ch.~Weinheimer, M.~Schrader, J.~Bonn, Th.~Loeken, H.~Backe, 
                 Nucl. Instrum. Methods A {\bf 311}, 273 (1993) 

\bibitem{Ullrich} Holger Ulrich, Dipl. Thesis, Mainz, Univ. (2000)

\bibitem{Schall} Jean-Pierre Schall, Dipl. Thesis, Mainz, Univ. (2001)

\bibitem{Mueller} Beatrix  M\"uller, Dipl. Thesis, Mainz, Univ. (2002)

\bibitem{Schwamm} Frank Schwamm, Doct. Thesis, Karlsruhe, Univ. (2004)

\bibitem{Thuem02} Thomas Th\"ummler, Dipl. Thesis, Mainz, Univ. (2002) 

\bibitem{Sanchez} Salvadore Sanchez, Dipl. Thesis, Mainz Univ. (2003)

\bibitem{flatt04} Bj\"orn Flatt et al., {\it to be published}

\bibitem{bonn99} J.~Bonn, L.~Bornschein, B.~Degen, E.W.~Otten, 
                 Ch.~Weinheimer, NIM {\bf A421}, 256 (1999) 

\bibitem{cibo} J.~Ciborowski, J.~Rembielinski, Eur. Phys. C {\bf 8}, 157 (1999)

\bibitem{Barthetal98} H.~Barth, L.~Bornschein, B.~Degen, L.~Fleischmann, 
         M.~Przyrembel, H.~Backe, A.~Bleile, J.~Bonn, D.~Goldmann, 
         M.~Gundlach, O.~Kettig, E.W.~Otten, G.~Tietze, Ch.~Weinheimer, 
         P.~Leiderer, O.~Kazachenko, A.~Kovalik, 
         Prog. Part. Nucl. Phys. {\bf 40}, 353 (1998)
 
\bibitem{auflad} B.~Bornschein, J.~Bonn, L.~Bornschein, E.W.~Otten, 
                 Ch.~Weinheimer, Journal of Low Temperature Physics, 
                 {\bf 131}, 69 (2003)

\bibitem{Wilkerson} J.F.~Wilkerson et al., Phys. Rev. Lett. {\bf 58}, 2023 
                    (1987) 

\bibitem{saenz_private} A.~Saenz, Humbold Univ., Berlin, {\it priv. commun.}

\bibitem{lobashev01} V.M.~Lobashev, talk and proc. Int. Conf. on 
                     Non-Accelerator New Physics, Dubna, 2001

\bibitem{lobashev03} V.M.~Lobashev, Proc. 17. Int. Conf. on  Nuclear Physics 
                     in Astrophysics, Debrecen/Hungary, 2002, Nucl. Phys. 
                     {\bf A719} (2003) 153c-160c 

\bibitem{losala} R.G.H.~Robertson et al., Phys. Rev. Lett. {\bf 87}, 957 (1991)

\bibitem{zurich} E.~Holzschuh et al., Phys. Lett. {\bf B287}, 381 (1992)

\bibitem{tokyo} H.~Kawakami et al., Phys. Lett. {\bf B256}, 105 (1991)

\bibitem{beijing} H.C.~Sun et al., CJNP {\bf 15}, 261 (1993)

\bibitem{liver} W.~Stoeffl, D.J.~Decman, Phys. Rev. Lett. {\bf 75}, 3237 (1995)

\bibitem{unified} G.J.~Feldmann and R.D.~Cousins, Phys. Rev. {\bf D57}, 3873 
                  (1998)

\bibitem{Lobashev98} V.M.~Lobashev, Prog. Port, Nucl. Phys. {\bf 40}, 
                     337 (1998)


\end{thebibliography}
\end{document}